\documentclass[letter,bibyear]{aa} % for the letters

\usepackage{epsfig}
\usepackage{amsmath}
\usepackage{subfigure}
\usepackage{natbib}
\usepackage{multirow}
\usepackage{color}
\usepackage{longtable}
\usepackage{graphicx}
\usepackage{epstopdf}
\usepackage{booktabs}
\usepackage{txfonts}
\newcommand{\jms}{J.~Mol.~Spectrosc.}

\newcommand{\jmst}{J.~Mol.~Struct.}

\newcommand{\kms}{km s$^{-1}$}

\begin{document}

\title{Discovery of two cyano derivatives of acenaphthylene (C$_{12}$H$_8$)
in TMC-1 with the QUIJOTE line survey\thanks{Based on
observations carried out
with the Yebes 40m telescope (projects 19A003,
20A014, 20D023, 21A011, 21D005, and 23A024). The 40m
radio telescope at Yebes Observatory is operated by the Spanish Geographic Institute
(IGN, Ministerio de Transportes, Movilidad y Agenda Urbana).}}

\author{
J.~Cernicharo\inst{1},
C.~Cabezas\inst{1},
R.~Fuentetaja\inst{1},
M.~Ag\'undez\inst{1},
B.~Tercero\inst{2,3},
J.~Janeiro\inst{4},
M.~Juanes\inst{5},
R.~I.~Kaiser\inst{6},
Y.~Endo\inst{7},
A.~L.~Steber\inst{5},
D.~P\'erez\inst{4},
C.~P\'erez\inst{5},
A.~Lesarri\inst{5},
N.~Marcelino\inst{2,3},
P.~de~Vicente\inst{2}
}

\institute{Dept. de Astrof\'isica Molecular, Instituto de F\'isica Fundamental (IFF-CSIC),
C/ Serrano 121, 28006 Madrid, Spain. \newline \email jose.cernicharo@csic.es
\and Centro de Desarrollos Tecnol\'ogicos, Observatorio de Yebes (IGN), 19141 Yebes, Guadalajara, Spain.
\and Observatorio Astron\'omico Nacional (OAN, IGN), C/ Alfonso XII, 3, 28014, Madrid, Spain.
\and Centro Singular de Investigaci\'on en Qu\'imica Biol\'oxica e Materiais Moleculares (CiQUS) and Departamento de Qu\'imica Org\'anica, Universidade de Santiago de Compostela, 15782 Santiago de Compostela, Spain
\and Departamento de Qu\'imica Física y Qu\'imica Inorg\'anica, Facultad de Ciencias-I.U. CINQUIMA, Universidad de Valladolid, 47011 Valladolid, Spain
\and Department of Chemistry, University of Hawaii at Manoa, Honolulu, HI 96822, USA
\and Department of Applied Chemistry, Science Building II, National Yang Ming Chiao Tung University, 1001 Ta-Hsueh Rd., Hsinchu 300098, Taiwan
}

\date{Received 10 September 2024;accepted 18 September 2024}

\abstract{We report the discovery in TMC-1 of two cyano derivatives of the
PAH acenaphthylene (C$_{12}$H$_8$). We have found two series of lines with the QUIJOTE line survey that we assign to 1-C$_{12}$H$_7$CN and 5-C$_{12}$H$_7$CN.
For the 1-isomer, we have detected and assigned 173 rotational transitions with $J$ up to 46 and $K_a$ up to 9, corresponding to 107 independent frequencies. For the 5-isomer, the identification is based on 56  individual lines,
corresponding to 117 rotational transitions with $J$ up to 40 and $K_a$ up to 8.
Identification of the carriers was achieved through a careful analysis of the derived rotational constants, which permit us to focus on molecules larger than naphthalene but smaller than anthracene and phenanthrene. Moreover, the derived rotational constants indicate that the species are planar; this allows us to discard derivatives of fluorene and acenaphthene, which are non-planar species. Quantum chemical calculations and subsequent chemical synthesis of these
molecules, as well as the observation of their rotational transitions in the laboratory, unequivocally support our identifications.
%Both isomers of cyanoacenapthylene, together with some other PAHs, present a peculiarity for their $a$ and $b$
%type transitions with $K_a$=0,1. It consists in that the frequencies of these transitions are near harmonically
%related but with half integer quantum numbers. The derived effective rotational constant,
%$B_{eff}$, is very close to the rotational constant $C$ obtained from a fit to all observed lines using a Watson Hamiltonian.
We also confirm, via a robust  line-by-line detection, the previous claimed detection of
1- and 2- cyanonaphthalene, which were obtained through statistical stacking techniques.
The column densities of 1- and 5-cyanoacenaphthylene are (9.5\,$\pm$\,0.9)$\,\times$\,10$^{11}$ cm$^{-2}$, while
those of 1- and 2-cyanonapthalene are (5.5\,$\pm$\,0.5)$\,\times$\,10$^{11}$ cm$^{-2}$. Hence, it seems that
acenaphthylene could be a factor of 1.7 more abundant than naphthalene. These results support a scenario in which PAHs grow in cold dark clouds based on fused five- and six-membered carbon  rings.}


\titlerunning{Cyanoacenaphthylene in TMC-1}
\authorrunning{Cernicharo et al.}

\maketitle

\section{Introduction}
The QUIJOTE\footnote{\textbf{Q}-band \textbf{U}ltrasensitive \textbf{I}nspection \textbf{J}ourney
to the \textbf{O}bscure \textbf{T}MC-1 \textbf{E}nvironment} ultrasensitive line survey \citep{Cernicharo2021a}
has provided the detection of more than 60 molecular species in the last four years. Some of them are very polar
molecules with relatively low abundances such as the 16 S-bearing species discovered in TMC-1 \citep{Cernicharo2024},
the radicals H$_2$C$_3$N and H$_2$C$_4$N \citep{Cabezas2021,Cabezas2023},
and the double-cyanide
derivatives of methane (NCCH$_2$CN) and ethylene (NCCHCHCN) \citep{Agundez2024}.
Nevertheless, one of the most significant results of QUIJOTE is the discovery, through the standard method of line-by-line
detection, of low-dipole but very abundant pure hydrocarbons,
such as CH$_2$CHCCH \citep{Cernicharo2021b}, $o$-C$_6$H$_4$ \citep{Cernicharo2021a}, CH$_2$CCHCCH \citep{Cernicharo2021c},
$c$-C$_5$H$_6$, and $c$-C$_9$H$_8$ \citep[cyclopentadiene and indene;][]{Cernicharo2021d}, the radical H$_2$CCCH \citep{Agundez2021,Agundez2022}, the long carbon chain CH$_2$CCHC$_4$H \citep{Fuentetaja2022}, and fulvenallene
\citep[$c$-C$_5$H$_5$CCH$_2$;][]{Cernicharo2022}.
These neutrals, together with some cations, such as $l$-C$_3$H$_3$$^+$ \citep{Silva2023}, certainly play an important role in the growth of larger hydrocarbons and PAHs in TMC-1 \citep{Cernicharo2022}.

Indene is the only polycyclic aromatic hydrocarbon (PAH) discovered directly from its
rotational spectrum \citep{Cernicharo2021d}. However, the cyano derivatives of benzene and naphthalene have been reported
in TMC-1 using stacking techniques with the GOTHAM line survey \citep{McGuire2018,McGuire2021}. It is not clear yet how these PAHs are formed in a cold dark cloud.
It is unlikely that these molecules arise from a reservoir of PAHs existing since the early stages of the cloud, because these relatively
small PAHs would not have survived the diffuse cloud stage. Moreover, the spatial distribution of benzonitrile is identical to that of cyanopolyynes and other molecular
species \citep{Cernicharo2023a}. Hence, the observational evidence points towards a bottom-up formation process for
these species in TMC-1.
The chemical ingredients that form PAHs are the carbon chains, radicals, and cations that have been detected with QUIJOTE and mentioned above.
However, the detailed chemical routes for the growth of PAHs are only beginning to emerge \citep{Kaiser2021,Cernicharo2022}. Taking into account the large abundances of indene and those inferred for benzene
and naphthalene from their cyanide derivatives, it is clear that
just a few reactions are involved in
their production. Unfortunately, these reactions are still missing in our chemical networks.
Consequently, it is highly important to detect more PAHs in order to provide additional clues as to their chemistry.

In this work, we present the discovery from the QUIJOTE data of two cyano derivatives of
acenaphthylene ($c$-C$_{12}$H$_8$), which is composed of three fused carbon rings
(two six-membered and one five-membered). We also confirm the previous identification of cyanonaphthalenes \citep{McGuire2021}  through a robust
line-by-line detection.
The derived abundances
indicate that cyanoacenaphthylenes are $\sim$1.7 more abundant than cyanonaphthalenes.
This result suggests that bigger PAHs in TMC-1
probably consist of a mixture of  fused five- and six-membered carbon rings.
We discuss possible formation routes for these molecules in cold clouds, such as TMC-1.

\section{Observations}
The observational data presented in this work are part of the QUIJOTE spectral line survey
\citep{Cernicharo2021a} in the Q-band towards TMC-1(CP) ($\alpha_{J2000}=4^{\rm h} 41^{\rm  m}
41.9^{\rm s}$ and $\delta_{J2000}=+25^\circ 41' 27.0''$), which was performed at the Yebes 40m radio
telescope. This survey was carried out using a receiver built within the Nanocosmos
project\footnote{\texttt{https://nanocosmos.iff.csic.es/}} consisting of two cooled high-electron-mobility-transistor (HEMT) amplifiers covering the 31.0-50.3 GHz band with horizontal
and vertical polarization. Fast Fourier transform spectrometers (FFTSs) with $8\times2.5$ GHz
and a spectral resolution of 38.15 kHz provide the whole coverage of the Q-band in both polarizations.
%The receiver temperatures achieved in the first observations vary from 22 K at 32 GHz to 42 K at 50 GHz.
%Some power adaptation in the down-conversion chains have reduced the
Receiver temperatures are 16\,K at 32 GHz and 30\,K at 50 GHz.
The experimental setup is described in detail by \citet{Tercero2021}.

All observations were performed using frequency-switching observing mode with a frequency throw of 10 and 8 MHz.
The total observing time on the source for data taken with frequency throws of 10 MHz and 8 MHz is 465 and 737 hours,
respectively. Hence, the total observing time on source is 1202 hours.
The QUIJOTE sensitivity varies between 0.08 mK at 32 GHz and 0.2 mK at 49.5 GHz.
A detailed description of the line survey and the data-analysis procedure are
provided in \citet{Cernicharo2021a,Cernicharo2022}.
The main beam efficiency
can be given across the Q band as $B_{\rm eff}$=0.797 exp[$-$($\nu$(GHz)/71.1)$^2$]. The forward telescope efficiency
is 0.97. The telescope beam size at half power intensity is 54.4$''$ at 32.4 GHz and
36.4$''$ at 48.4 GHz. The absolute calibration uncertainty is 10\,$\%$.
The data were analysed with the GILDAS package\footnote{\texttt{http://www.iram.fr/IRAMFR/GILDAS}}.

\section{Results}\label{results}
Line identification in this work was performed using the
MADEX code \citep{Cernicharo2012}, in conjunction with the CDMS
and JPL catalogues \citep{Muller2005,Pickett1998}. The intensity scale used in this study is the antenna temperature ($T_A^*$). Consequently, the telescope parameters and source properties were used when modelling the emission of the different species in order to produce synthetic spectra on this temperature scale. In this work, we assumed a velocity for the source relative to the local standard at rest of 5.83 \kms\, \citep{Cernicharo2020}. The source was assumed to be circular with a uniform brightness temperature and a radius of 40$''$ \citep{Fosse2001}.

\subsection{A peculiar way to search for PAHs}
We would like to start by introducing the method we used to search for new PAHs.
Although seemingly disconnected, the story deserves to be told. When we detected
the anion C$_7$N$^-$ \citep{Cernicharo2023b}, we searched intensively for C$_7$N.
Quantum chemical calculations by \citet{Botschwina1999} predicted an electronic ground state $^2\Pi$
with a moderate dipole moment of 1\,D. However, due to the possible admixing
of the ground state with a low-lying $^2\Sigma$ state, which has
a dipole moment of $\sim$3.6\,D \citep{Botschwina1999}, the true
value of the dipole moment could be between those of the two states,
and in this case it could be detectable with
our QUIJOTE data. The rotational
constant of C$_7$N is predicted to be $\sim$\,585 MHz. If the electronic ground
state is $^2\Pi$, then the quantum numbers should be half integers. Our automatic
programs to search for harmonic relations indicate 12 lines in very good harmonic
relation with $J$ half integers and $B_{\rm eff}$\,=\,580.987 MHz and $D_{\rm eff}$\,=\,4.94 Hz.
Hereafter, we refer to these half-integer quantum numbers as $J^*$.
This was
a very exciting result, pointing towards the presence of C$_7$N in TMC-1. However,
all these harmonically related lines could be easily identified with the MADEX catalogue
with $K_a$\,=\,0, 1, $J_u$\,=\,$J^*-1/2$ transitions of 1-cyanonaphthalene
(see Table \ref{fit_1-cna}).
These lines are shown in Fig. \ref{fig_1-cna},
which is based on a modified Loomis-Wood diagram \citep{Loomis1928}. Details of the confirmation
of cyanonaphthalenes and of the harmonic relation with $J$ half integers
for $a$- and $b$-type $K_a$\,=\,0, 1 transitions are given in Appendix\,\ref{confirmation_cnas}.

We verified that this peculiarity of the $K_a$\,=\,0, 1 lines also applies ---with some variations due
to the value of the rotational constants and to the values of the dipole moment components--- to many other PAHs
such as indene, fluorene, cyanoantracene, and cyanophenantrene among others.
For any oblate asymmetric rotor, pairing of lines will occur for high-$J$; first
for $K_c$=$J$, and then for $K_c$=$J$-1 and so on, and the effective $B$ will approach $C$.
A detailed analysis of the asymptotic behaviour of the energy levels of an asymmetric
rotor is provided by \citet{Watson2007}. Consequently,
visual inspection of the data
following the method described in Appendix \ref{confirmation_cnas} allows us to easily detect the rotational
transitions of heavy PAHs.

\subsection{The series of lines B429 and B444}\label{detec_Bs}
We searched using automatic and/or visual procedures for series of lines in harmonic relation with integer or half-integer quantum numbers. Once the series were
found,
we used the method described in the previous section and in Appendix \ref{confirmation_cnas}
to visually explore the
QUIJOTE data, searching for spectral patterns similar to those of 1- and 2-cyanonaphthalene.
We discovered two series of lines harmonically
related with half-integer quantum numbers. We quickly verified
that these series could not be assigned to a species with quantum numbers
2\,$J^*$ as all lines with even $J$
 would be missing. Hence, either we are detecting species in a $^2\Pi$ state, or we
are observing a similar case to that of 1- and 2-cyanonaphthalene.

\begin{figure}[]
\centering
\includegraphics[width=0.49\textwidth]{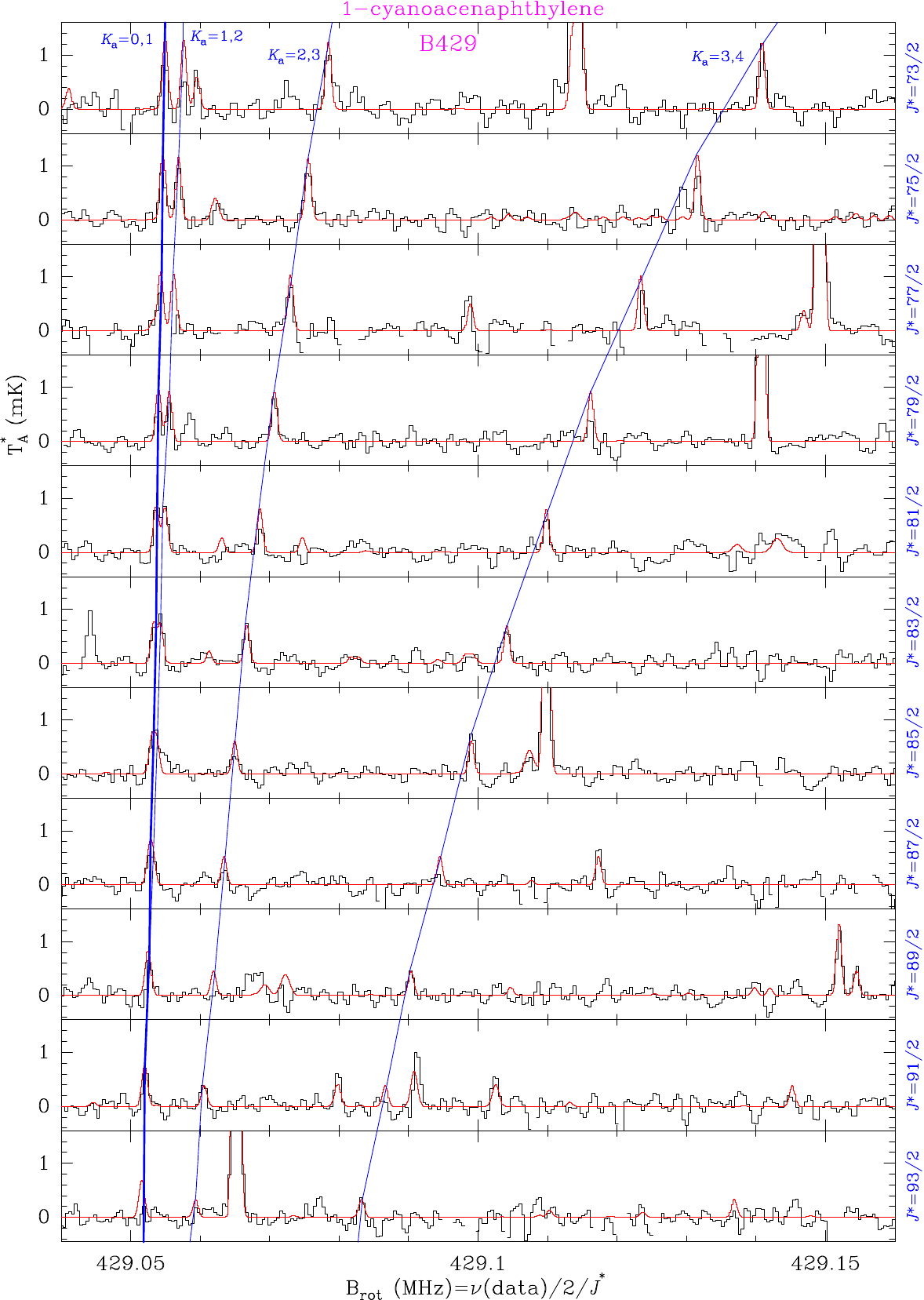}
\caption{Modified Loomis-Wood diagram of some of the observed lines of B429 (1-cyanoacenaphthylene).
The abscissa corresponds to the value of the rotational constant, which has been fixed in this plot
to values of between 429.040 and 429.160 MHz. The ordinate is the antenna temperature
---corrected for atmospheric and telescope losses--- in millikelvin. Each box presents the QUIJOTE
data for frequencies 2\,$B_{rot}$\,$J^*$, where $J^*=J_u+1/2$.
The red line corresponds to the synthetic spectrum computed for TMC-1, including the rotational
transitions of all molecular species detected with QUIJOTE, adopting for them the
physical parameters derived in all previous works. This spectrum contains the rotational lines of the
two isomers of cyanoacenaphtylene discovered in this paper. The rotational transitions of 1-cyanoacenaphthylene correspond to those connected through blue lines (see text). The blue line on the left side corresponds to the $K_a$\,=\,0, 1 transitions that are harmonically related with half-integer quantum numbers.}
\label{fig_B429}
\end{figure}

\begin{figure}[]
\centering
\includegraphics[width=0.49\textwidth]{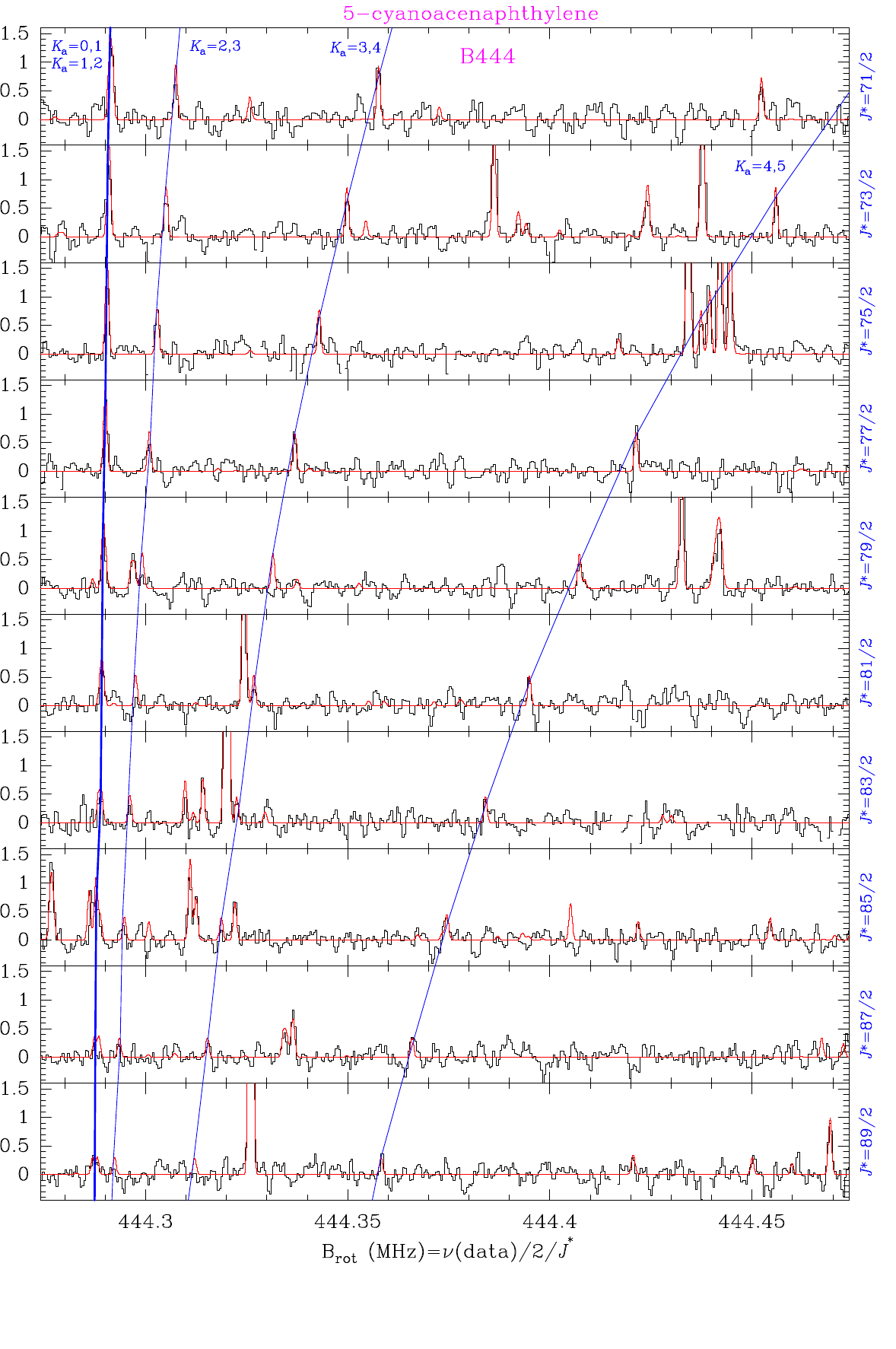}
\caption{Similar to Fig. \ref{fig_B429} but for B444 (5-cyanoacenaphthylene). The range of
$B_{rot}$ is from 444.274 to 444.474 MHz.}
\label{fig_B444}
\end{figure}

The first series could be fitted with $B_{\rm eff}$\,=\,429.0596\,$\pm$\,0.0003 MHz and $D_{\rm eff}$\,=\,1.81$\pm$0.07 Hz,
with $J^*$ from 73/2 to 93/2. The standard deviation of the fit is 11.6 kHz.
Hereafter, we refer to this series of lines as B429.
Using the same kind
of modified Loomis-Wood plot described above, we found additional series of lines
at slightly higher frequencies than those of B429.
The first four series of lines, including those of B429, are shown in Fig.
\ref{fig_B429}. To facilitate their visual identification, they are connected by blue straight lines, similarly to the visualization of 1-cyanonaphthalene shown in Fig. \ref{fig_1-cna}. The lines connected by the first blue
line (left one) correspond to upper quantum numbers $J_u$,
$K_a$\,=\,0, 1, and $K_c=J_u$, where $J_u$ for each panel is $J^*-1/2$ ($J^*$ is indicated at the right side
of the panels).
These lines are those of B429 and are perfectly reproduced with the constants $B_{\rm eff}$ and $D_{\rm eff}$ derived
above.
The second blue line connects the transitions $J_u-1$, $K_a$\,=\,1, 2, and
$K_c=J_u-2$; it collapses to the first blue line for $J^*$\,=\,83/2 ($J_u$\,=\,41).
The third blue line connects the transitions $J_u-2$, $K_a$\,=\,2, 3, and
$K_c=J_u-3$. All transitions are $\Delta J$\,=\,+1.
Many additional series of lines
related to B429 were found in our data involving up to a total of 107 independent frequencies.
The derived line parameters are given in Table \ref{line_parameters}.
It is clear that all these
lines belong to the progressions of an asymmetric molecule
with integer quantum numbers $J_u=J^*-1/2$
($J_u$ from 35 up to 46).

In summary, the lines of B429 belong to $K_a$\,=\,0, 1 (the set of transitions connected by the
first blue line). The other lines have values of $K_a$ varying from 1 up to 9. It was relatively
easy to assign the quantum numbers to the 107 observed lines (corresponding to 173 rotational transitions) and to derive from these frequencies the rotational constants given in Table \ref{main_rot_constants}.
We used an $A$-reduced Hamiltonian (representation $I^r$) and the derived rotational
constants are given in Table \ref{main_rot_constants}. Similar to the case of 1- and
2-cyanonaphthalene, the rotational constant $C$ derived differs from $B_{\rm eff}$ by
less than 0.002 MHz.

The second series of lines in harmonic relation with half-integer quantum numbers is shown in Fig. \ref{fig_B444}. The lines are connected by a straight vertical blue line and can be reproduced  with $B_{\rm eff}$\,=\,444.298526\,$\pm$\,0.00024 MHz and $D_{\rm eff}$\,=\,2.848\,$\pm$\,0.077 Hz, with
$J^*$ between 71/2 and 85/2. Hereafter, we refer to the carrier of these lines as B444. Additional series of lines are shown in Fig. \ref{fig_B444}; these are also connected by blue lines that are assigned to $K_a$\,=\,1, 2, and 3 lines of an asymmetric rotor, similarly to B429. The line parameters of all the observed lines are given in Table \ref{line_parameters}. A total of 56 independent lines corresponding to 117 rotational transitions were assigned to B444. The rotational and distortion constants of B444 derived using an $A$-reduced Hamiltonian (representation $I^r$) are given in Table \ref{main_rot_constants}.

For both carriers, the $a$- and $b$-type transitions have a significant collapse for $K_a$\,=\,0, 1 and 1, 2. However, if the $b$ component of the dipole moment is large, then we can expect to detect individual $b$-type transitions, as in the case of 1- and 2- cyanonaphthalene (see Fig. \ref{fig_1-cna_2}). We searched for these $b$-type transitions for B429 and B444 without success, which indicates that the rotational spectrum is dominated by $a$-type transitions, that is, that $\mu_b$ is significantly lower than $\mu_a$.

\subsection{Quantum chemical calculations and identification of 1- and 5-cyanoacenaphthylene}\label{ab_initio}
In order to search for the carriers of B429 and B444, we compared the derived rotational constants with those of other asymmetric species to get some insights in their structure. The derived rotational constants for B429 and B444 could be compatible
with cyano (or CCH) derivatives of anthracene, phenantrene, acenaphtene, fluorene, and acenapthylene, among other  species with three fused rings.

We can discard derivatives of anthracene and phenanthrene as the rotational constants of
these species, before any derivation, are smaller than those of B429 and
B444 \citep{Baba2009,Kowaka2012}. To bolster our confidence in this conclusion, we
performed quantum chemical calculations (see Appendix \ref{app_ab_initio}) for all cyano
derivatives of anthracene (see Table \ref{constants_3rings}) and phenanthrene (see
Table \ref{cnp}). None of these species, which are depicted in Fig. \ref{fig_antra}
and \ref{fig_phenan}, can fit our rotational constants.

\begin{table*}
\caption{Molecular constants of the cyano isomers of acenaphthylene derived from QUIJOTE data (representation $A$ $I^r$.} \label{main_rot_constants}
\centering
\begin{tabular}{|l|cc|c|c|cc|}
\hline
Parameter          & 1-CNACY$^a$& B429 (TMC-1)  & 3-CNACY$^a$          & 4-CNACY$^a$  & 5-CNACY$^a$&B444 (TMC-1)  \\
\hline
$A$ (MHz)          &   1271.64  & 1272.1707(19) &     1476.89          &   1451.83    & 1248.93    &1246.5874(75)  \\
$B$ (MHz)          &    647.47  & 647.27938(19) &      608.98          &    565.62    &  688.66    & 690.1787(24)  \\
$C$ (MHz)          &    429.03  & 429.06155(14) &      431.18          &    407.04    &  443.90    & 444.29815(32) \\
$\Delta_J$ (Hz)    &            &  2.079(41)    &                      &              &            &   2.15(10)    \\
$\Delta_{JK}$ (Hz) &            &  9.59(63)     &                      &              &            &  21.2(12)     \\
$\Delta_K$ (Hz)    &            &  249.3(95)    &                      &              &            & 195(16)       \\
\hline
$N_{trans}^b$, $N_{lines}^c$       &            & 173, 107    &                      &              &            & 117, 56           \\
%$N_{lines}^c$      &            & 107           &                      &              &            & 56            \\
$J_{max}$, $K_{a,max}$          &            & 46, 9            &                      &              &            & 42, 9            \\
%$K_{a,max}$         &            &  9            &                      &              &            &  9            \\
$\sigma^d$(kHz),$\sigma_w^e$    &            & 10.5, 0.85          &                      &              &            &  9.7, 0.85          \\
%$\sigma_w^e$       &            & 0.85          &                      &              &            &  0.85         \\
$\mu_a$, $\mu_b$  (D)        &   5.6, 0.1      &               &     5.1, 1.1              &   5.3, 0.1        & 4.6, 1.0        &               \\
%$\mu_b$ (D)        &   0.1      &               &     1.1              &   0.1       & 1.0        &               \\
\hline
\end{tabular}
\tablefoot{
\tablefoottext{a}{Theoretical values of the rotational constants (see Sect. \ref{ab_initio}).}
\tablefoottext{b}{Total number of rotational transitions.}
\tablefoottext{c}{Total number of independent frequencies.}
\tablefoottext{d}{Standard root mean square deviation of the fit in kHz.}
\tablefoottext{e}{Weighted root mean square deviation of the fit.}
}
\end{table*}

 We can also discard non-planar molecules, as the derived $C$ rotational constant for B429
 and B444 is almost identical to $(A$ $\times$ $B)/(A+B)$, that is, the derived
 inertial defect is small or compatible with that of cyano derivatives of planar PAHs  \citep[see
 e.g.][]{McNaughton2018}. This allows us to discard cyano or ethylenic derivatives of indene,
 fluorene, and acenaphthene as possible carriers of our lines, as these species have large
 inertial defects due to the presence of one or two pairs of hydrogen atoms out of the
 plane of the molecule. Fluorene and acenaphthene have been observed in the laboratory
 by \citet{Thorwirth2007}, and the cyano derivatives of fluorene have also been observed \citep{Cabezas2024}. Rotational constants for these species, together with values obtained from our quantum chemical calculations for CN derivatives of
 acenaphthene,
 are given in Tables\,\ref{cnf} and \ref{cnacenaphthene}. Indene derivatives have been
 studied in detail in the laboratory by \citet{Sita2022}, and their rotational constants
 do not match those of B429 and B444. Doubly cyano derivatives of naphthalene, which are
 depicted in Fig. \ref{dcn_fig}, can also be discarded as carriers of our lines as their
 rotational constants are far from those of B429 and B444 (see Appendix  \ref{app_ab_initio}
 and Table \ref{dcn}).

\begin{figure}[]
\centering
\includegraphics[width=0.46\textwidth]{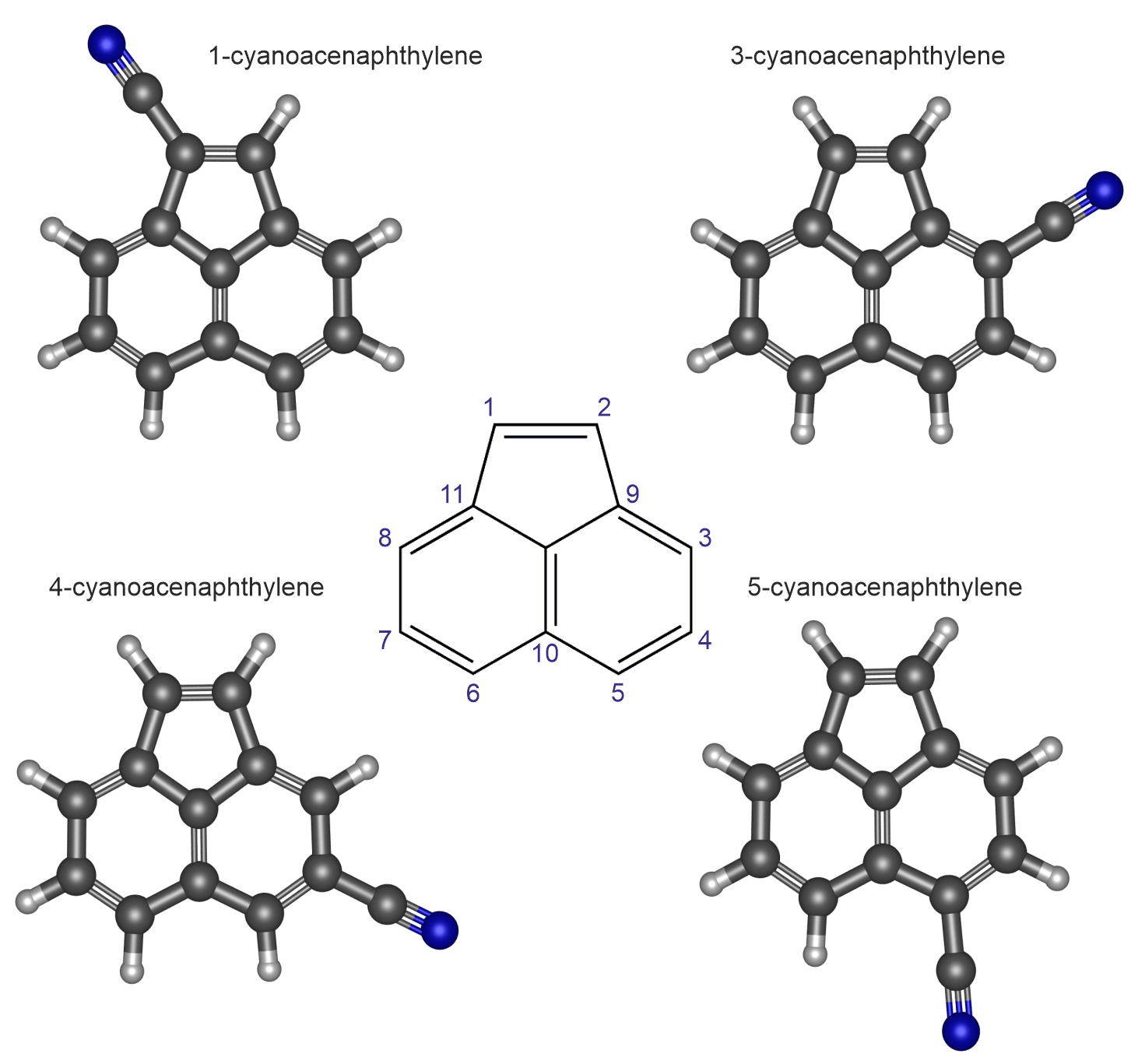}
\caption{Chemical structure of acenaphthylene (center design) and its cyano derivatives. The numbers label the nitrile substitution positions of the four possible isomers of
cyanoacenaphthylene. The isomers detected in this work are 1-CNACY and 5-CNACY, which correspond to B429 and B444, respectively.}
\label{fig_isomers}
\end{figure}

 Finally, we consider the cyano derivatives of acenaphthylene (C$_{12}$H$_8$), a PAH that has
 been studied in the laboratory by \citet{Thorwirth2007} and that is planar (see central
 structure in Fig. \ref{fig_isomers}).
 The rotational constants derived in the laboratory for this species are compared to those
 derived through our quantum chemical calculations in Table \ref{acn_rot_constants}.
 Four possible cyano derivatives can be obtained for this species. The structure of the
 mother molecule and of its CN derivatives is shown in Fig. \ref{fig_isomers}. Hereafter, we
refer to the CN derivatives of acenaphthylene as n-CNACY (n=1, 3, 4 or 5). Our quantum
 chemical calculations indicate that 1-CNACY can be identified with B429 and that B444 can
 be assigned to 5-CNACY. The predicted rotational constants for the four isomers of CNACY
 (n-C$_{12}$H$_7$CN; n=1, 3, 4, and 5) are given in Table \ref{main_rot_constants}.

 In order to fully confirm the identification of B429 and B444 with 1-CNACY and 5-CNACY,
 we synthesized them in the chemical laboratory (see Appendix \ref{app_synthesis}) and
 observed their rotational spectrum in the microwave laboratory (see Appendix
 \ref{app_FTMW}). The resulting molecular parameters are given in Tables \ref{1cnace}
 and \ref{5cnace}. The rotational constants derived in space for B429 agree with those
 of 1-CNACY (see Table \ref{1cnace}), and those of B444 are in perfect agreement with those
 of 5-CNACY (see Table \ref{5cnace}). Hence, we firmly conclude that we have detected, for the first time in space, two of the cyano derivatives of the PAH
 acenaphthylene. This is the fist time that a 3-fused carbon ring is detected in space.
 Moreover, these are the first PAHs that are fully characterised from astronomical data
 prior to any laboratory information on its rotational spectrum. These molecules contain
 21 atoms and are, apart from the fullerenes, the largest species found in space.
  We find no evidence in our data for the 3-CNACY and 4-CNACY; moreover, they
 cannot be produced in the laboratory following the procedure described in Appendix \ref{app_synthesis}.

\section{Discussion}\label{Discussion}
%\subsection{Column densities}
In order to derive column densities for the two cyanide derivatives of acenaphthylene, we assumed that these species
are close to thermalisation, $T$$_{rot}$= 9\,K, that molecular emission
follows that of C$_6$H$_5$CN \citep{Cernicharo2023a}, and that it can be simplified, as in all our previous
analyses, as a source of uniform brightness temperature of 80$''$ in diameter.
With these assumptions, we derive the same
column density of (9.5$\pm$0.5)$\times$10$^{11}$ cm$^{-2}$ for 1-CNACY and 5-CNACY.

For 1- and 2-cyanonaphthalene we derive, adopting the same assumptions, a common column density of (5.5$\pm$0.5)$\times$10$^{11}$
cm$^{-2}$. These values are a factor 1.4 times smaller than those derived by \citet{McGuire2021} for the
same species. However, our identification
of 1- and 2-cyanonaphthalene is based on many individual lines and a rotational temperature of 9\,K,
while those of \citet{McGuire2021} result from a
stacking of spectral data and the assumption of four velocity components with different spatial sizes.
We note that
our extremely sensitive map of benzonitrile \citep{Cernicharo2023a} does not support the spatial sizes derived
from the stacking procedure.
Any velocity component present in the stacked data of \citet{McGuire2021} should
be present in our integrated intensity  map, which indicates that benzonitrile has the same spatial distribution
as many other species in TMC-1.

Assuming that the mother
species, acenaphthylene and naphthalene, are ten times more abundant than their cyano derivatives, we could expect column
densities for acenaphthylene ($c$-C$_{12}$H$_8$) and naphthalene
($c$-C$_{10}$H$_8$) of $\sim$10$^{13}$ cm$^{-2}$  and $\sim$5.5$\times$10$^{12}$ cm$^{-2}$ , respectively.

Acenaphthylene itself has a small permanent dipole moment ($\mu_b$=0.31 D) and has been observed in the
laboratory \citep{Thorwirth2007}. Only a 3$\sigma$ upper limit of N$\le$3$\times$10$^{13}$ cm$^{-2}$ can be obtained from the QUIJOTE.
This result is in line with the abundance derived from its cyano derivatives. Acenaphthene ($c$-C$_{12}$H$_{10}$, $\mu_b$=0.9\,D)
and fluorene ($c$-C$_{13}$H$_{10}$, $\mu_b$=0.53\,D) have also been observed by \citet{Thorwirth2007}. Unfortunately, only upper limits of  $\sim$(5-10)$\times$10$^{12}$ cm$^{-2}$ can be obtained for their
column densities.

The estimated column density of acenaphthylene is larger than that of naphthalene by a factor of
1.7, which is similar to the column density ratio between indene and acenaphthylene, of namely N=(1.6$\pm$0.3)$\times$10$^{13}$ cm$^{-2}$
 \citep{Cernicharo2021d}.
Hence, it seems that the growth of PAHs in TMC-1 favors the
presence of PAH species containing fused five- and six-membered carbon rings. It is important
to note in this context that the column density of cyclopentadiene
is also significantly large, N=(1.2$\pm$0.3)$\times$10$^{13}$ cm$^{-2}$
\citep{Cernicharo2021d}, which is of the order of the
column density that can be estimated for benzene from benzonitrile \citep{Cernicharo2022}.
Current chemical networks fail to explain the abundances of the simplest cycles, such
as benzene and cyclopentadiene \citep{Cernicharo2022}. These latter authors examined formation
routes to $c$-C$_5$H$_6$ involving
cations and found that invoking the reaction between $l$-C$_3$H$_3$$^+$ and C$_2$H$_4$
---which measurements suggest is rapid
and was found to form the cation C$_5$H$_7$$^+$ as a product \citep{Smyth1982,Anicich2003}--- enabled production of the
observed abundance of cyclopentadiene. That is, both neutral-neutral
and ion-neutral reactions are likely to be at the origin of the growth of PAHs in cold
clouds such as TMC-1. The formation of acenaphthylene itself could involve smaller rings,
such as indene and naphthalene, as precursors. For example, the reaction of C$_2$H with
naphthalene or that of C$_3$H with indene could directly produce acenaphthylene, although
this possibility still has to be verified theoretically or/and experimentally (see Appendix
\ref{chemical_routes}).

\begin{acknowledgements}
We thank Spanish Ministerio de Ciencia e Innovaci\'on and the European Regional Development Fund (MICINN–ERDF) for funding support through projects PID2019-106110GB-I00, PID2019-106235GB-I00, PID2021-125015NB-I00, PID2022-139933NB-I00, and PID2023-147545NB-I00.
We also thank ERC for funding through grant ERC-2013-Syg-610256-NANOCOSMOS. We thank the CSIC (Spain) for funding through project PIE 202250I097. The present study was also supported by Ministry of Science and Technology of Taiwan and CSIC under the MoST-CSIC Mobility Action 2021 (Grants 11-2927-I-A49-502 and OSTW200006).
%JJ thank the Agencia Estatal de Investigación for the award of a pre-doctoral fellowship (PRE2020-092897)
%DP and JJ thank financial support from the Xunta de Galicia (Centro de investigación do Sistema universitario de Galicia
%accreditation 2023-2027, ED431G 2023/03) and the European Union (European Regional Development Fund - ERDF)

\end{acknowledgements}

\normalsize
\begin{appendix}
\section{Confirmation of the presence of cyanonaphthalenes in TMC-1}\label{confirmation_cnas}
\begin{table}
\tiny
\caption{Fit of the $K_a$=0 lines of 1-cyanonaphthalene in the Q-band.} \label{fit_1-cna}
\centering
\begin{tabular}{ccccc}
\hline
Transition\,$^a$& $J^*$\,$^b$& $\nu_{\rm calc}$\,$^c$&  $\nu_{\rm lin}$\,$^d$& $\nu_{\rm calc}-\nu_{\rm lin}$\,$^e$\\
               &                     & (MHz)  & (MHz)           &   (kHz)\\
\hline
$27_{0,27}-26_{0,26}$ & 55/2   & 31953.900 & 31953.897     &   3.0\\
$28_{0,28}-27_{0,27}$ & 57/2   & 33115.827 & 33115.826     &   1.0\\
$29_{0,29}-28_{0,28}$ & 59/2   & 34277.751 & 34277.751     &   0.0\\
$30_{0,30}-29_{0,29}$ & 61/2   & 35439.672 & 35439.672     &   0.0\\
$31_{0,31}-30_{0,30}$ & 63/2   & 36601.589 & 36601.590     &  $-$1.0\\
$32_{0,32}-31_{0,31}$ & 65/2   & 37763.503 & 37763.504     &  $-$1.0\\
$33_{0,33}-32_{0,32}$ & 67/2   & 38925.413 & 38925.414     &  $-$1.0\\
$34_{0,34}-33_{0,33}$ & 69/2   & 40087.320 & 40087.321     &  $-$1.0\\
$35_{0,35}-34_{0,34}$ & 71/2   & 41249.222 & 41249.223     &  $-$1.0\\
$36_{0,36}-35_{0,35}$ & 73/2   & 42411.120 & 42411.121     &  $-$1.0\\
$37_{0,37}-36_{0,36}$ & 75/2   & 43573.014 & 43573.015     &  $-$1.0\\
$38_{0,38}-37_{0,37}$ & 77/2   & 44734.904 & 44734.904     &   0.0\\
$39_{0,39}-38_{0,38}$ & 79/2   & 45896.789 & 45896.789     &   0.0\\
$40_{0,40}-39_{0,39}$ & 81/2   & 47058.669 & 47058.669     &   0.0\\
$41_{0,41}-40_{0,40}$ & 83/2   & 48220.545 & 48220.544     &   1.0\\
$42_{0,42}-41_{0,41}$ & 85/2   & 49382.416 & 49382.415     &   1.0\\
\hline
\end{tabular}
\tablefoot{
\tablefoottext{a}{Rotational quantum numbers of $a$-type $K_a$\,=\,0 transitions of
1-cyanonaphthalene with $K_c$\,=\,$J_u$.}
\tablefoottext{b}{Effective rotational quantum number ($J^*=J_u+1/2$).}
\tablefoottext{c}{Predicted frequencies using an $A$-reduced Hamiltonian (representation $I^r$) and the rotational constants
determined by \citet{McNaughton2018}.}
\tablefoottext{d}{Fitted frequencies using the relation $\nu$\,=\,2\,$B_{\rm eff}$\,$J^*$\,$-$\,4\,$D_{\rm eff}$\,$J^*$$^3$. The results of the fit are
$B_{\rm eff}$\,=\,580.987413\,$\pm$\,0.000015 MHz and $D_{\rm eff}$\,=\,4.9354\,$\pm$\,0.0057 Hz. The standard deviation of the fit is 1 kHz.
}
\tablefoottext{e}{Difference between the predicted frequencies using an $A$-reduced Hamiltonian (representation $I^r$)
and those obtained through the linear fit with half integer quantum numbers.}
}
\end{table}
\normalsize

\begin{figure}[]
\centering
\includegraphics[width=0.49\textwidth]{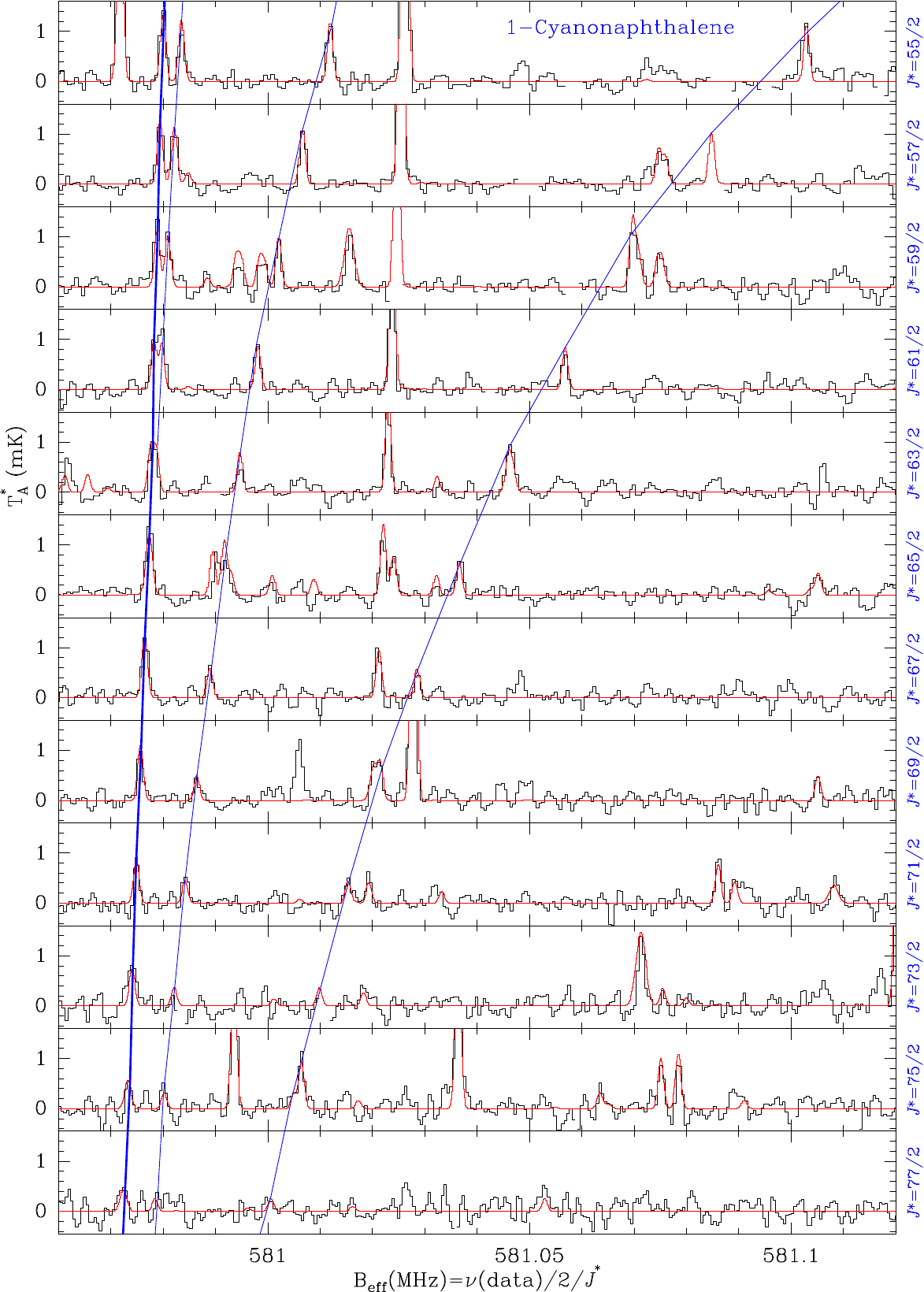}
\caption{Modified Loomis-Wood diagram of some of the observed lines of 1-cyanonaphthalene.
The abscissa corresponds to the value of the rotational constant. The ordinate is the antenna temperature,
corrected for atmospheric and telescope losses, in milli Kelvin. Each box presents the QUIJOTE
data for frequencies 2\,$B_{rot}$\,$J^*$, where $J^*=J_u+1/2$.
The red line corresponds to the synthetic spectrum computed for TMC-1 including the rotational
transitions of all molecular species detected with QUIJOTE adopting for them the
physical parameters derived in previous works. Features without a red line counterpart are unknown lines.
The model contains the rotational lines of the
two isomers of cyanonaphthalene and cyanoacenaphtylene reported in this work.
The rotational transitions
of 1-cyanonaphthalene correspond to the lines connected through blue lines (see text). The left blue line corresponds to the transitions harmonically related with half integer quantum numbers ($K_a$\,=\,0, 1; $a$ and $b$-type transitions; $J_u$=$J^*$-1/2). The other
blue lines connect transitions of 1-cyanonaphthalene corresponding to $K_a$=1, 2, 3 and 4 ($a$ and $b$-type).}
\label{fig_1-cna}
\end{figure}

\begin{figure*}[]
\centering
\includegraphics[width=0.8\textwidth]{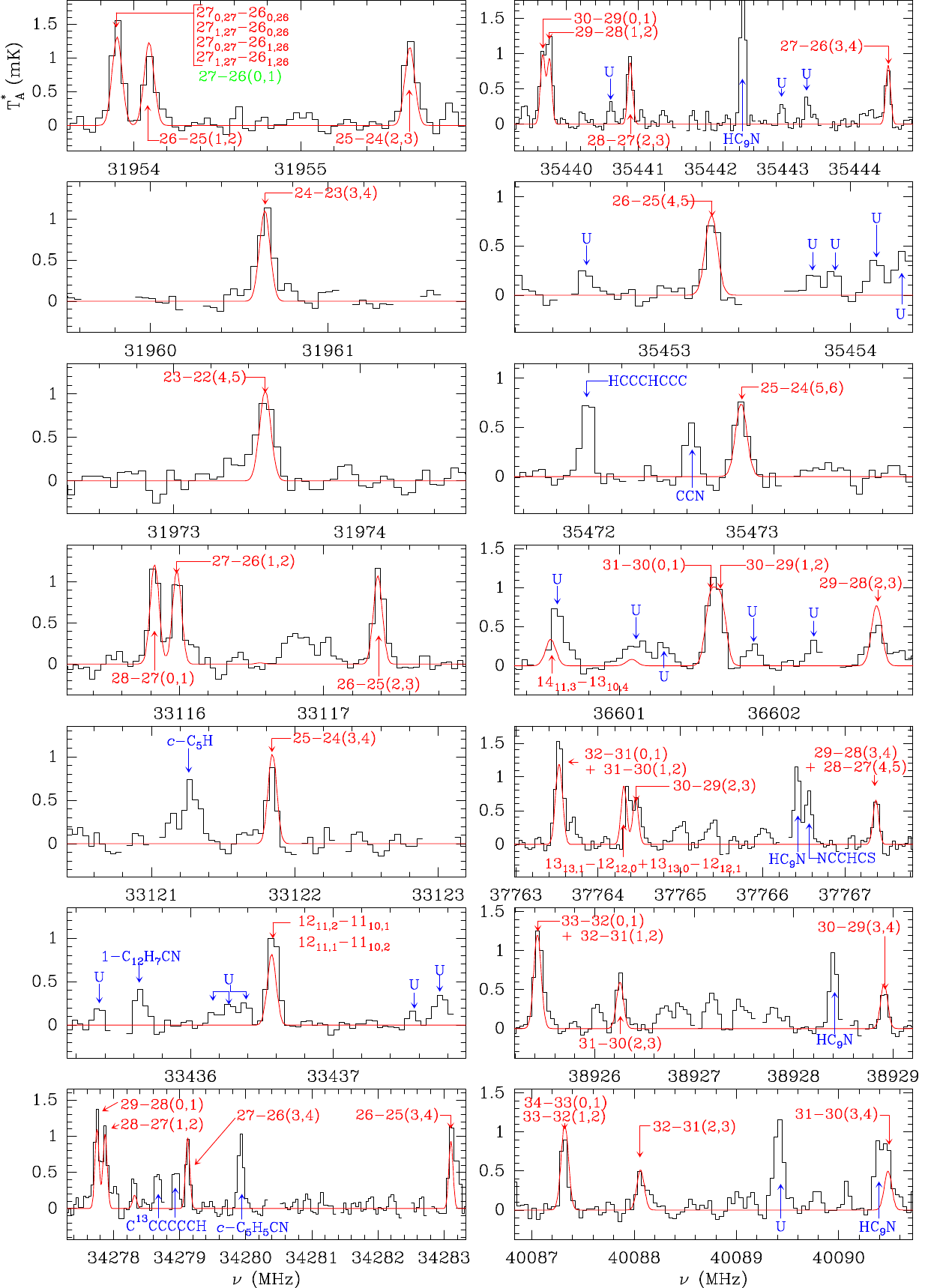}
\caption{Some of the lines of Fig. \ref{fig_1-cna} over enhanced frequency scales. The lines are indicated by arrows and their quantum numbers
are depicted for each feature. The red line corresponds to the synthetic spectra derived from the model presented in Sect. \ref{Discussion}. Other lines from molecules previously studied or  unknown features are indicated in blue.}
\label{fig_1-cna_2}
\end{figure*}

As noted in Sect. \ref{results}, we were looking for harmonic relations with half
integer quantum numbers that
could be related to C$_7$N. A dozen lines were found that, as explained in Sect.
\ref{results}, could be fitted with a simple relation
$\nu$\,=\,2\,$B_{\rm eff}$\,$J^*$\,$-$\,4\,$D_{\rm eff}$\,$J^*$$^3$. These lines are
shown in Fig. \ref{fig_1-cna} (lines connected with the blue line at the left). Each
panel of this figure
represents the data of the survey with frequencies 2\,$B_{rot}$\,$J^*$, with
$J^*$ being $J_u+1/2$ and $J_u$ the integer upper quantum number of the transitions
$K_a$\,=\,0, 1 of 1-cyanonaphthalene. This species has $a$- and $b$-type transitions,
with similar
values of the dipole moment \citep{McNaughton2018}, and has
the peculiarity that the $a$-type $J'_{0,J'}-J''_{0,J''}$ and $J'_{1,J'}-J''_{1,J''}$
with $J''=J'-1$
collapse to the same frequency than the $b$-type $J'_{1,J'}-J''_{0,J''}$ and
$J'_{0,J'}-J''_{1,J''}$.
Hence, the four $K_a=0,1$ $a$- and $b$-lines are at the same frequency which produces
a natural stacking
favouring the detection of these lines. These four transitions with $K_a$=0,1 will be
denoted as $J'-J''$(0,1).
Examples of them are given in Fig. \ref{fig_1-cna_2} where the quantum numbers $J'$
and $J''$
are indicated. All lines in this figure labelled as $J'-J''$(n,n+1) correspond to two
$a$- and two $b$-type collapsed
transitions with upper and lower rotational quantum numbers $J'$ and $J''$,
respectively.

The Ray asymmetry parameter \citep{Ray1932}, $\kappa=(2B-A-C)/(A-C)$, of 1-cyanonaphthalene
is around $-$0.1, hence, the molecule is highly asymmetric.
Nevertheless, a clear series of lines harmonically related
appear in the diagram of Fig. \ref{fig_1-cna} as connected by a vertical line slightly tilted due to the contribution
of the rotational distortion to the frequencies of the transitions. The calculated frequencies of these transitions using the rotational and distortion constants derived by
\citet{McNaughton2018} are given in Table \ref{fit_1-cna}, together with the fitted
frequencies assuming that the lines arise from a linear
molecule with half integer quantum numbers. The match
between the frequencies calculated using an $A$-reduced Hamiltonian (representation $I^r$)
for an asymmetric molecule and those obtained from a Hamiltonian for a linear molecule with half integer quantum numbers ($J^*$=$J_u$+1/2)
is simply astonishing.

The values of $B_{\rm eff}$ and $D_{\rm eff}$ derived through the linear fit are 580.987413\,$\pm$\,0.000015 MHz, and
4.9354\,$\pm$\,0.0057 Hz, respectively. The parameter $B_{\rm eff}$ agrees extremely
well with the rotational constant $C$ derived by \citet[][$C$\,=\,580.988976\,$\pm$\,0.000006 MHz]{McNaughton2018}.
It should be pointed out that the $K_a$\,=\,0, 1 lines cannot be
reproduced with a fit to a linear species using integer quantum numbers.
The additional series of lines that appear
in the figure correspond to $K_a$ values of 1, 2, 3, and 4 of $a$ and $b$ type. They appear
connected by significantly curved blue lines due to the high contributions of non-diagonal terms of the
Hamiltonian and of the distortion terms of the molecule.
In Fig. \ref{fig_1-cna_2} we show a zoom for several of these lines to illustrate that
they are clearly detected on a line-by-line basis.
The behaviour of the $K_a=0,1$ lines of 1-cyanonaphthalene is the expected result for
high-$J$ values of large molecules \citep{Watson2007}.
In this case, the lines of an asymmetric top in the
oblate asymptotic limit ($J$=$K_c$), or
the prolate asymptotic limit ($J$=$K_a$), collapse to a single
feature producing strong lines separated by 2$C$ or 2$A$ that appear
as progressions in the spectrum. Up to four lines can be collapsed to the
same frequency (see, e.g., Fig. \ref{fig_1-cna_2}).
%The observed behaviour for
%the frequencies of 1-cyanonaphthalene in the Q-band can
%be reproduced if E($J$)=$C\times$$J(J+1)$+$C\times$$J$.
%, i.e.,
%$\nu(J$$\rightarrow$$J-1)$=2$\times$$C$$\times$$J$+C= 2$\times$$C$$\times$($J$+1/2) =
%2$\times$$C$$\times$$J^*$.

\begin{figure}[]
\centering
\includegraphics[width=0.49\textwidth]{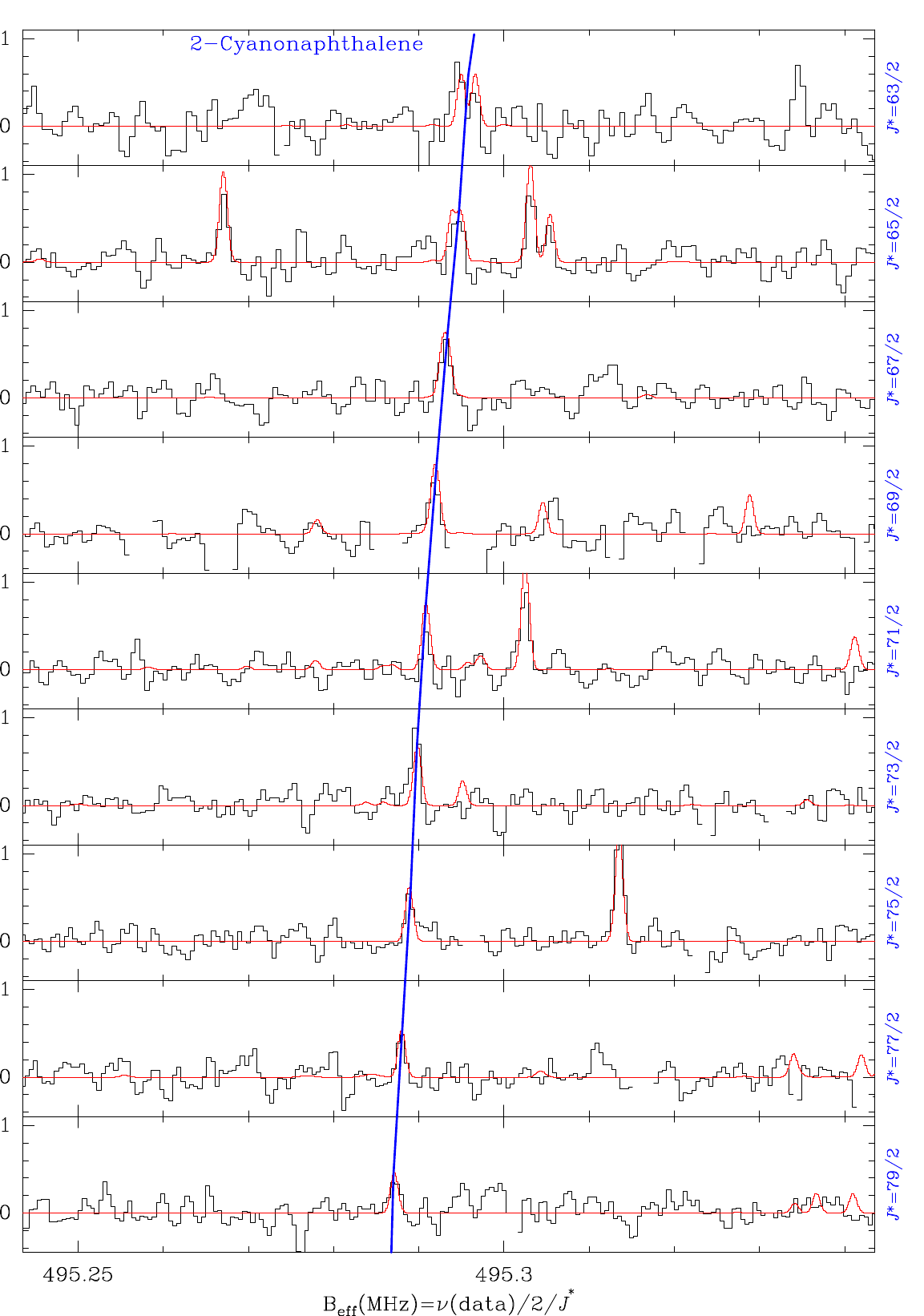}
\caption{Modified Loomis-Wood diagram for the observed $K_a$\,=\,0,1 lines of 2-cyanonaphthalene.
The abscissa corresponds to the value of the rotational constant. The ordinate is the antenna temperature,
corrected for atmospheric and telescope losses, in milli Kelvin. Each box presents the QUIJOTE
data for frequencies 2\,$B_{rot}$\,$J^*$, where $J^*=J_u+1/2$.
The red line corresponds to the synthetic spectrum computed for TMC-1 including the rotational
transitions of all molecular species detected in previous works and the two
isomers of cyanonaphthalene and cyanoacenaphthylene.
The rotational transitions
of 2-cyanonaphthalene correspond to the lines connected through
the blue vertical line (see text).
For this isomer of cyanonaphthalene they
correspond to the $a$-type transitions harmonically related with half integer
quantum numbers, $J^*$, and corresponding to $J_u=J^+-1/2$ and $K_a$\,=\,0, 1.
}
\label{fig_2-cna}
\end{figure}

\begin{figure}[]
\centering
\includegraphics[width=0.48\textwidth]{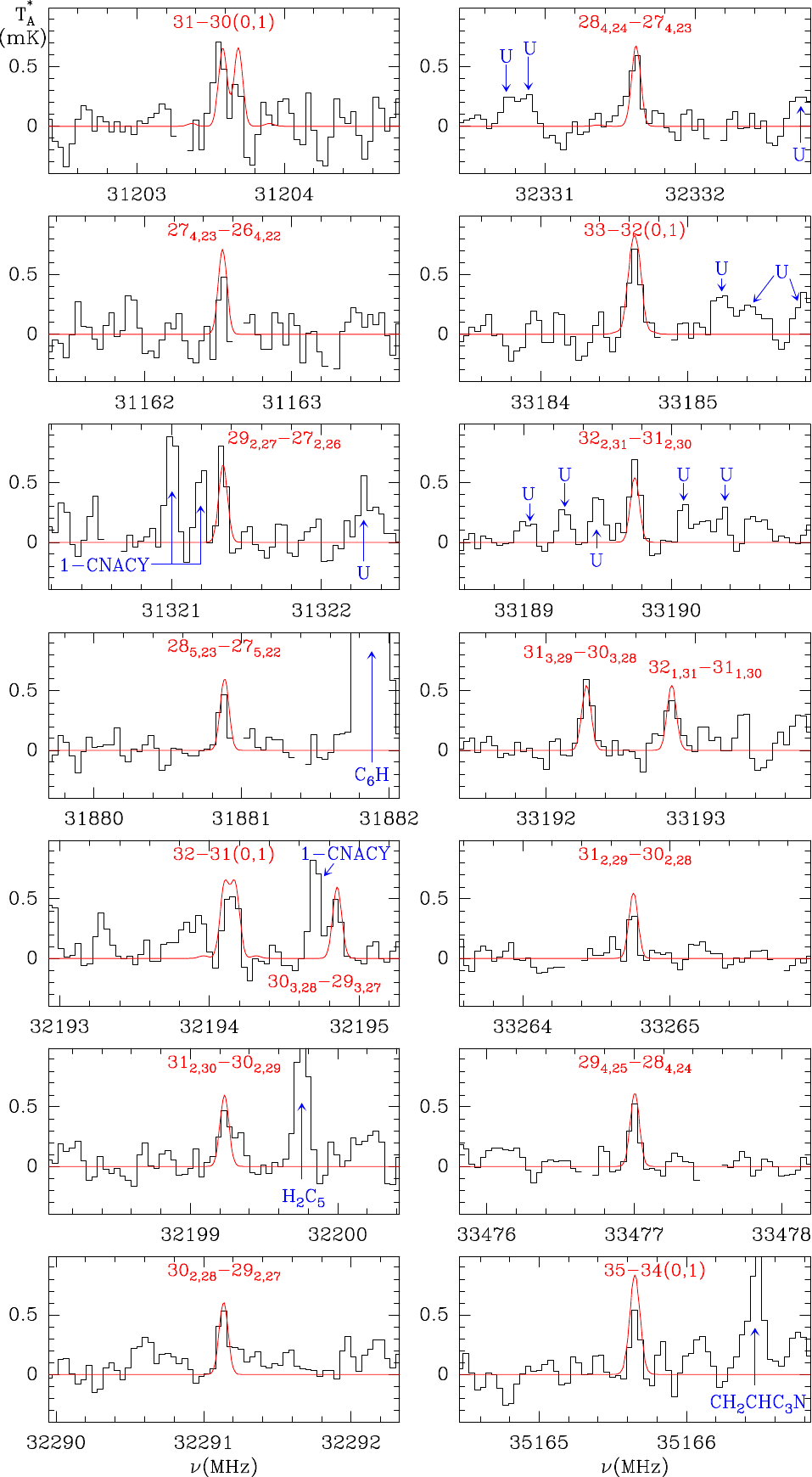}
\caption{Additional lines of 2-cyanonaphthalene (see Fig. \ref{fig_2-cna}). The quantum numbers
are indicated for each feature. The red line corresponds to the synthetic spectra of the model for 2-cyanonaphthalene presented in Sect. \ref{Discussion}.
Other lines from molecules previously studied or  unknown features are indicated in blue.}
\label{fig_2-cna_2}
\end{figure}

The lines of the isomer 2-cyanonaphthalene are shown in Fig. \ref{fig_2-cna}.
For this isomer the Ray's asymmetry parameter, $\kappa$, is $-$0.9, which means that the species is near
the prolate symmetric limit ($\kappa$\,=\,$-$1). For this isomer, the spectrum is dominated by $a$-type transitions, which are 25 times stronger than the $b$-type ones ($\mu_a$\,=\,5.2\,D, $\mu_b$\,=\,1.0\,D; \citealt{McNaughton2018}). For $J'$\,=\,31 the two $K_a$\,=\,0 and $K_a$\,=\,1 lines are separated by 100 kHz
(see top panel of Fig. \ref{fig_2-cna}),
while for $J'$\,=\,35 the splitting of the two lines is only of 30 kHz and appear unresolved in our data
(see panel corresponding to $J^*$\,=\,71/2 in Fig. \ref{fig_2-cna}). The $a$-type $J'-J''$(0,1)
lines can be reproduced with $B_{\rm eff}$\,=\,495.317 MHz, $D_{\rm eff}$\,=\,13.2 Hz, and $H_{\rm eff}$\,=\,1.53 $\mu$Hz with
a standard deviation of 19 kHz. A better result (5.8 kHz) is obtained by fitting an additional distortion term.
The effective rotational constant using half integer  quantum numbers for these $K_a$\,=\,0, 1 transitions
is very similar to the $C$ rotational constant ($C$\,=\,495.29352 MHz) derived for this isomer by \citet{McNaughton2018}.
Additional lines with higher $K_a$ values are shown in Fig. \ref{fig_2-cna_2}.

\section{Line parameters of the two CNACY isomers}
Line parameters for all observed transitions with the Yebes 40m radio telescope
were derived by fitting a Gaussian line profile to them
using the GILDAS package. A
velocity range of $\pm$20\,\kms\, around each feature was considered for the fit after a polynomial
baseline was removed. Negative features produced in the folding of the frequency switching data were blanked
before baseline removal. A view of some of the transitions of the two isomers of CNACY is given in figures \ref{fig_B429} and
\ref{fig_B444}.
%Additional lines of each species are shown in \textcolor{red}{Fig. XX2}.

\onecolumn
\begin{longtable}{lcrcccc}
\caption{Observed line parameters towards TMC-1 for the two isomers of CNACY$^\$$.}
\label{line_parameters}\\
\hline \hline
Transition& $\nu_{obs}$$^a$  & Obs-Cal$^b$& $\int$ $T_A^*$ dv $^c$ & $\Delta$v$^d$   &   T$_A^*$$^e$& Notes\\
          &      (MHz)       & (kHz)      & (mK\,km\, s$^{-1}$)    &  (km s$^{-1}$)  &   (mK)       & \\
\endfirsthead
\caption{continued.}\\
\hline \hline
Transition& $\nu_{obs}$$^a$  & Obs-Cal$^b$& $\int$ $T_A^*$ dv $^c$ & $\Delta$v$^d$   &   T$_A^*$$^e$& Notes\\
          &      (MHz)       & (kHz)      & (mK\,km\, s$^{-1}$)    &  (km s$^{-1}$)  &   (mK)       & \\
\hline
\endhead
\hline
\endfoot
\hline
\textbf{1-CNACY}\\
 $36_{0,36}-35_{0,35}$  & 31321.003$\pm$0.010& $-$7 &0.77$\pm$0.07 & 0.70$\pm$0.07&  1.02$\pm$0.08& \\
 $36_{1,36}-35_{1,35}$  & 31321.003$\pm$0.010& $-$7 &              &              &               & \\
 $35_{1,34}-34_{1,33}$  & 31321.200$\pm$0.010& $-$7 &0.49$\pm$0.07 & 0.61$\pm$0.13&  0.76$\pm$0.08& \\
 $35_{2,34}-34_{2,33}$  & 31321.200$\pm$0.010& $-$7 &              &              &               & \\
 $34_{2,32}-33_{2,31}$  & 31322.721$\pm$0.010& $-$5 &0.81$\pm$0.07 & 0.79$\pm$0.11&  0.96$\pm$0.08& \\
 $34_{3,32}-33_{3,31}$  & 31322.721$\pm$0.010& $-$5 &              &              &               & \\
 $33_{3,30}-32_{3,29}$  & 31327.286$\pm$0.010&    2 &1.00$\pm$0.09 & 0.78$\pm$0.05&  1.20$\pm$0.06& \\
 $33_{4,30}-32_{4,29}$  & 31327.286$\pm$0.010&    2 &              &              &               & \\
 $32_{4,28}-31_{4,27}$  & 31338.050$\pm$0.010&    8 &0.90$\pm$0.07 & 0.72$\pm$0.07&  1.18$\pm$0.08& \\
 $32_{5,28}-31_{5,27}$  & 31338.050$\pm$0.010&    5 &              &              &               & \\
 $31_{6,26}-30_{6,25}$  & 31361.167$\pm$0.030&    6 &0.85$\pm$0.08 & 0.80$\pm$0.00&  0.99$\pm$0.12&A\\
 $31_{5,26}-30_{5,25}$  & 31361.263$\pm$0.030&$-$18 &0.37$\pm$0.08 & 0.80$\pm$0.00&  0.41$\pm$0.12&A\\
 $30_{7,24}-29_{7,23}$  & 31408.519$\pm$0.020&    9 &0.52$\pm$0.07 & 0.80$\pm$0.00&  0.60$\pm$0.10&A\\
 $30_{6,24}-29_{6,23}$  & 31412.542$\pm$0.010&   27 &0.76$\pm$0.11 & 1.09$\pm$0.10&  0.65$\pm$0.11& \\
 $28_{9,20}-27_{9,19}$  & 31424.864$\pm$0.010&   22 &0.74$\pm$0.09 & 0.90$\pm$0.12&  0.77$\pm$0.10& \\
 $27_{8,19}-26_{8,18}$  & 31478.333$\pm$0.010& $-$4 &0.68$\pm$0.07 & 0.96$\pm$0.15&  0.66$\pm$0.08& \\
 $29_{8,22}-28_{8,21}$  & 31483.445$\pm$0.020& $-$1 &0.55$\pm$0.10 & 0.96$\pm$0.18&  0.54$\pm$0.11& \\
 $29_{7,22}-28_{7,21}$  & 31564.045$\pm$0.020&    1 &0.29$\pm$0.08 & 0.76$\pm$0.28&  0.75$\pm$0.12& \\
 $28_{8,20}-27_{8,19}$  & 32162.846$\pm$0.010&    6 &0.72$\pm$0.12 & 0.82$\pm$0.10&  0.83$\pm$0.06& \\
 $37_{0,37}-36_{0,36}$  & 32179.095$\pm$0.010& $-$2 &0.94$\pm$0.10 & 0.90$\pm$0.11&  0.98$\pm$0.07& \\
\\
\\
\\
\hline
\hline
\textbf{5-CNACY}\\
 $35_{ 1,35}-34_{ 1,34}$& 31544.693$\pm$0.020&    29&1.74$\pm$0.07 & 1.14$\pm$0.05&  1.44$\pm$0.06& \\
 $34_{ 1,33}-33_{ 1,32}$& 31544.693$\pm$0.020& $-$21&              &              &               & \\
 $35_{ 0,35}-34_{ 0,34}$& 31544.693$\pm$0.020&    29&              &              &               & \\
 $34_{ 2,33}-33_{ 2,32}$& 31544.693$\pm$0.020& $-$21&              &              &               & \\
 $33_{ 3,31}-32_{ 3,30}$& 31545.822$\pm$0.010& $-$10&0.53$\pm$0.05 & 0.83$\pm$0.10&  0.60$\pm$0.06& \\
 $33_{ 2,31}-32_{ 2,30}$& 31545.822$\pm$0.010& $-$10&              &              &               & \\
 $32_{ 4,29}-31_{ 4,28}$& 31549.389$\pm$0.010&  $-$3&0.97$\pm$0.05 & 0.80$\pm$0.05&  1.14$\pm$0.06& \\
 $32_{ 3,29}-31_{ 3,28}$& 31549.389$\pm$0.010&  $-$3&              &              &               & \\
 $31_{ 5,27}-30_{ 5,26}$& 31557.882$\pm$0.010&  $-$9&0.83$\pm$0.12 & 0.93$\pm$0.16&  0.84$\pm$0.06&C\\
 $31_{ 4,27}-30_{ 4,26}$& 31557.882$\pm$0.010&  $-$9&              &              &               & \\
 $30_{ 6,25}-29_{ 6,24}$& 31576.097$\pm$0.010& $-$10&0.97$\pm$0.05 & 0.80$\pm$0.05&  0.96$\pm$0.09& \\
 $30_{ 5,25}-29_{ 5,24}$& 31576.097$\pm$0.010& $-$30&              &              &               & \\
 $29_{ 7,23}-28_{ 7,22}$& 31613.683$\pm$0.010& $-$11&0.31$\pm$0.05 & 0.62$\pm$0.15&  0.47$\pm$0.07& \\
 $29_{ 6,23}-28_{ 6,22}$& 31614.562$\pm$0.010&  $-$2&0.30$\pm$0.05 & 0.73$\pm$0.17&  0.38$\pm$0.07& \\
 $28_{ 8,21}-27_{ 8,20}$& 31685.056$\pm$0.010&    15&0.41$\pm$0.07 & 0.62$\pm$0.13&  0.63$\pm$0.09& \\
 $28_{ 7,21}-27_{ 7,20}$& 31708.838$\pm$0.010&  $-$9&0.21$\pm$0.07 & 0.83$\pm$0.26&  0.24$\pm$0.07& \\
 $36_{ 1,36}-35_{ 1,35}$& 32433.236$\pm$0.020&    11&0.80$\pm$0.08 & 0.80$\pm$0.00&  0.99$\pm$0.08&A\\
\hline
\end{longtable}
\tablefoot{\\
\tablefoottext{\$}{The full content of this table can be found in electronic
form at https://doi.org/10.5281/zenodo.13810127. All uncertainties correspond to 1$\sigma$.}\\
\tablefoottext{a}{Measured frequency assuming a v$_{LSR}$ of 5.83 km\,s$^{-1}$ for TMC-1 \citep{Cernicharo2020}.}\\
\tablefoottext{b}{Observed minus calculated frequencies in kHz. The calculated frequencies are those resulting from
the molecular constants obtained from a
fit of an $A$-reduced Hamiltonian (representation $I^r$) to the observed frequencies. The derived
rotational and distortion constants are
given in Table \ref{main_rot_constants}. Unresolved doublets are fitted to the averaged frequency of the
two transitions.}\\
\tablefoottext{c}{Integrated line intensity (in mK\,km\,s$^{-1}$).}\\
\tablefoottext{d}{Linewidth at half intensity derived by fitting a Gaussian function to
the observed line profile (in km\,s$^{-1}$).}\\
\tablefoottext{e}{Antenna temperature (in milli Kelvin).}\\
\tablefoottext{A}{Partially blended with another feature. A fixed linewidth of 0.80 km\,s$^{-1}$ has been adopted.}\\
\tablefoottext{B}{Only data from the observations with frequency throw of 8 MHz.}\\
\tablefoottext{C}{Only data from the observations with frequency throw of 10 MHz.}\\
\tablefoottext{D}{Line too broad. Probably blended with another feature. Parameters are uncertain.}\\
\tablefoottext{E}{Line probably blended with another feature. Parameters can be derived but are uncertain.}
\tablefoottext{F}{Narrow line. Parameters are uncertain}
}

\section{Quantum chemical calculations}\label{app_ab_initio}

Molecular geometry optimizations for all the plausible candidates were carried out using the density functional theory (DFT) variant B3LYP \citep{Becke1993} and the Pople triple-$\zeta$ basis set 6-311++G(d,p) \citep{Frisch1984}). From these calculations we obtained the values of the rotational constants and those of the electric dipole moment components. In addition, for some of the candidates we also performed harmonic frequency calculations at the same level of theory. All the calculations were done using the GAUSSIAN 16 package \citep{Frisch2016}. We chose this calculation method since it reproduces fairly well the rotational constants of the PAHs and their cyano derivatives at a reasonable computational cost, as shown in \citet{McNaughton2018}. See also as an example the comparative theoretical and experimental values obtained for acenaphthylene, shown in Table \ref{acn_rot_constants}.

We considered as plausible candidates the cyano derivatives of the PAHs with three fused rings: anthracene and phenanthrene ($c$-C$_{14}$H$_{10}$), $9H$-fluorene ($c$-C$_{13}$H$_{10}$), acenaphtene ($c$-C$_{12}$H$_{10}$), and acenaphtylene ($c$-C$_{12}$H$_{8}$). In addition, we also calculated the molecular constants for the dicyano derivatives of naphthalene.

\begin{figure}[]
\centering
\includegraphics[width=0.50\textwidth]{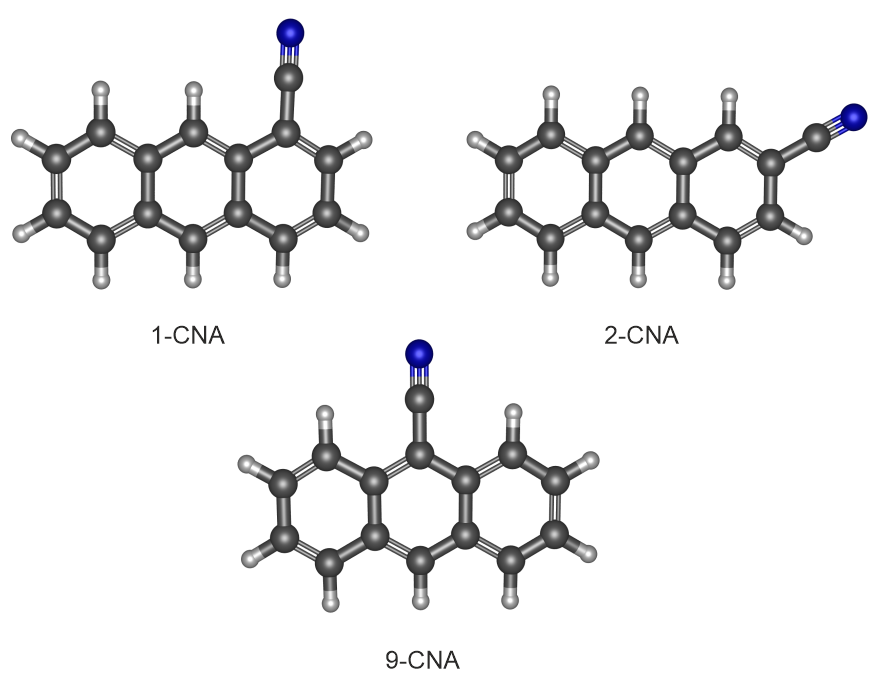}
\caption{Chemical structure of the CNA isomers.}
\label{fig_antra}
\end{figure}

\begin{figure}[]
\centering
\includegraphics[width=0.50\textwidth]{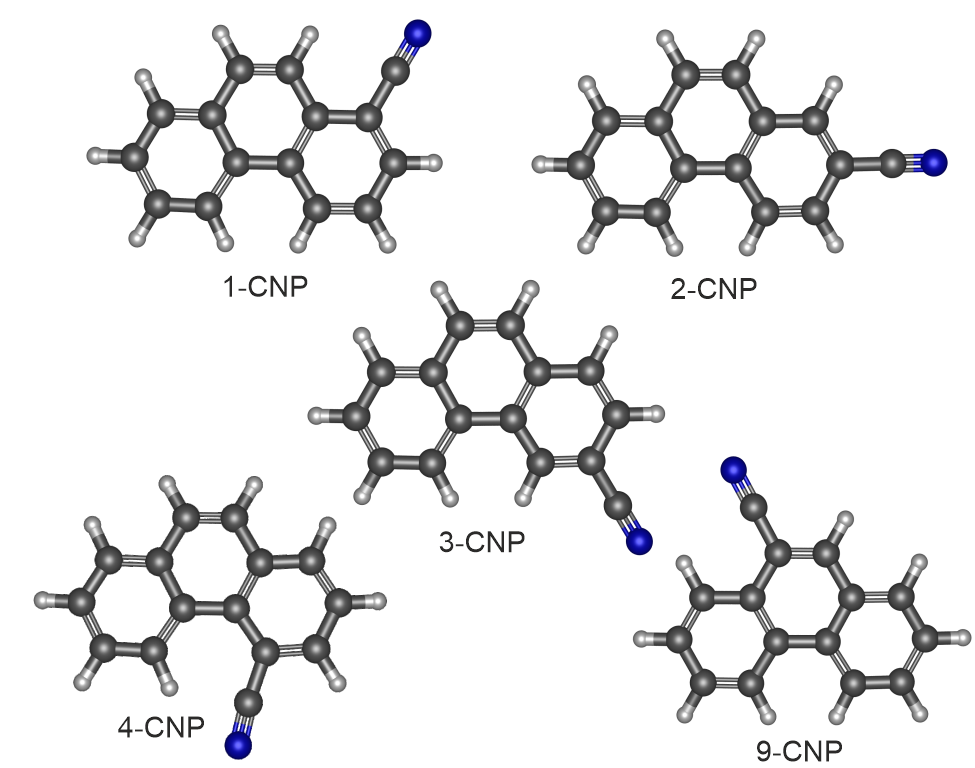}
\caption{Chemical structure of the CNP isomers.}
\label{fig_phenan}
\end{figure}

\begin{figure}[]
\centering
\includegraphics[width=0.45\textwidth]{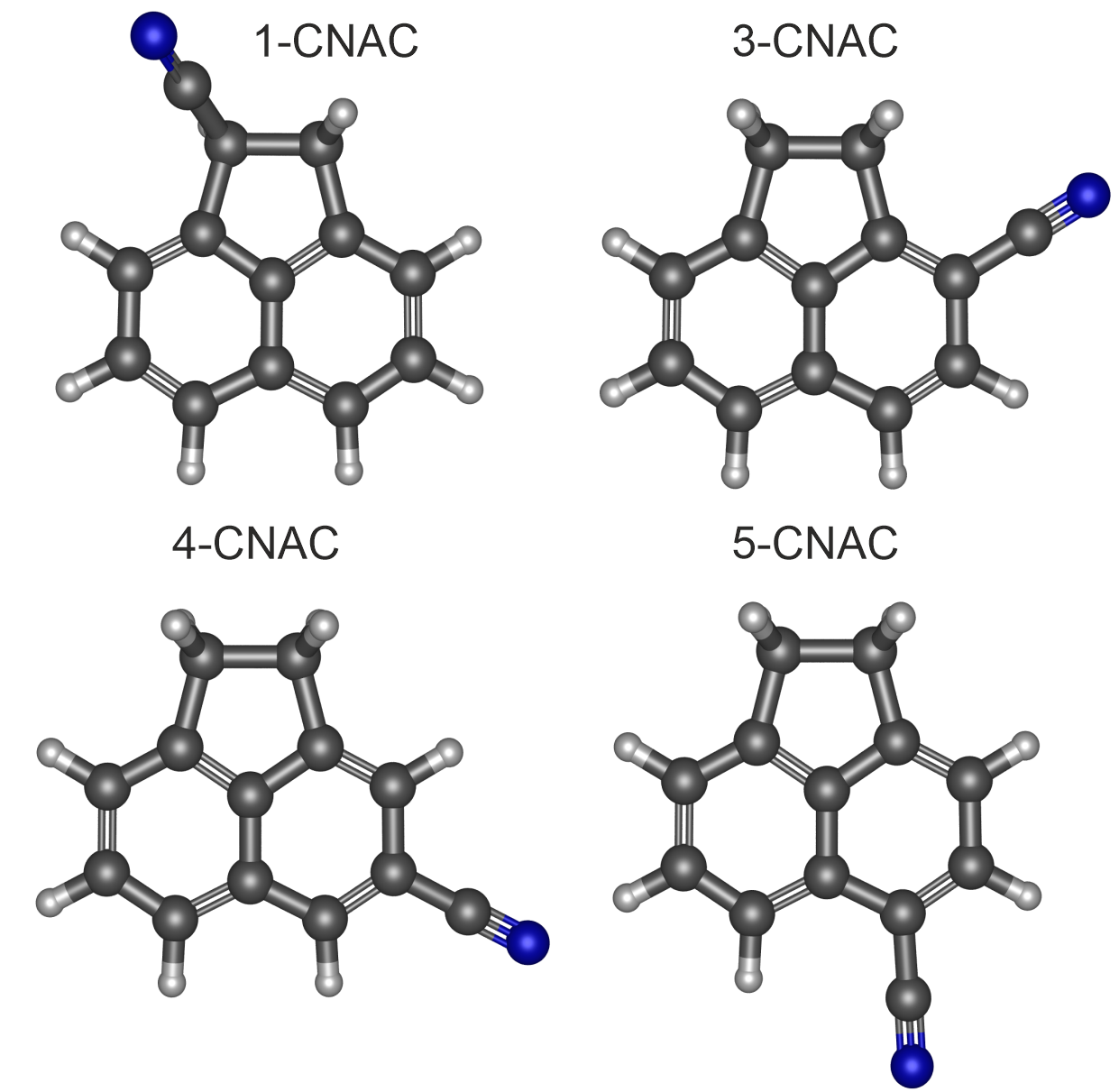}
\caption{Chemical structure of the CNAC isomers.}
\label{fig_acenaphthene}
\end{figure}

\begin{figure*}[]
\centering
\includegraphics[width=0.90\textwidth]{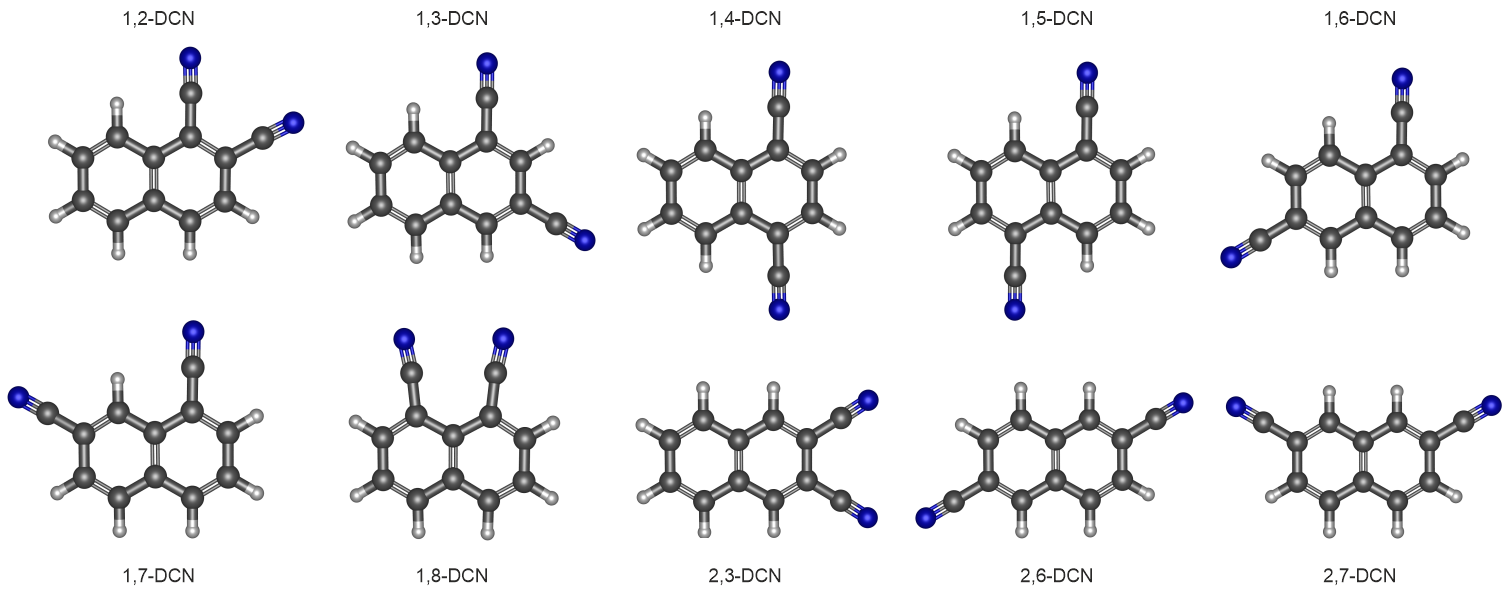}
\caption{Chemical structuresof the DCN isomers.}
\label{dcn_fig}
\end{figure*}

The molecular structures for all the isomers of cyanoanthracene (CNA) and cyanophenanthrene (CNP) are shown in Figs. \ref{fig_antra} and \ref{fig_phenan}, while the theoretical molecular constants are given in Table \ref{constants_3rings} and \ref{cnp}, together with those experimentally determined for the 9-cyanoanthracene (9-CNA) and 9-cyanophenanthrene (9-CNP) isomers, observed in the laboratory by \citet{McNaughton2018}. The molecular constants for the cyano derivatives of $9H$-fluorene (CNF) have been already published by \citet{Cabezas2024} together with the experimental constants, that are shown in Table \ref{cnf}. The calculated molecular constants for the five cyano derivatives of acenaphtene (CNAC) are shown in Table  \ref{cnacenaphthene} and their molecular structures in Fig. \ref{fig_acenaphthene} and those for the acenaphtylene are summarized in Table \ref{acn_rot_constants}. Finally the theoretical rotational constants for the ten different dicyano derivatives of naphthalene (DCN) are shown in Table \ref{dcn} and the molecular structures in Fig. \ref{dcn_fig}.

\begin{table}
\caption{Experimental and theoretical molecular constants of acenaphthylene.} \label{acn_rot_constants}
\centering
\begin{tabular}{lcc}
\hline
\hline
Parameter          & \citet{Thorwirth2007}   &  Theory$^a$     \\
\hline
$A$ (MHz)          &      1511.82609(12)     &     1513.0       \\
$B$ (MHz)          &      1220.63105(29)     &     1221.8       \\
$C$ (MHz)          &      675.529854(65)     &      675.9       \\
$\Delta_J$ (Hz)    &      31.4(26)           &      32.77       \\
$\Delta_{JK}$ (Hz) &      $-$49.2(68)          &     $-$57.51       \\
$\Delta_K$ (Hz)    &      20.8(44)           &      27.67       \\
$\delta_J$ (Hz)    &      $-$5.4(14)           &      4.05        \\
$\delta_K$ (Hz)    &      0.0                &     $-$12.52       \\
\hline
\hline
\end{tabular}
\tablefoot{
\tablefoottext{a}{This work.}
}
\end{table}

\begin{table*}
\caption{Molecular constants for the CNA isomers.} \label{constants_3rings}                                            \centering
\begin{tabular}{lccccc}
\hline
\hline
Parameter    & Anthracene$^a$ & 1-CNA  &  2-CNA  & 9-CNA  &  9-CNA$^b$ \\
\hline
$A$ (MHz)    &2146.24(12)  & 1091.3     &       1845.7    & 985.6          &  985.8526447(95) \\
$B$ (MHz)    &  452.540(74) & 386.2     &   279.2         & 450.6          &  451.2051876(206) \\
$C$ (MHz)    & 374.036(51)  & 285.2     &   242.5         & 309.2          & 309.6126249(178)     \\
$\mu_a$ (D)  & 0.0  &  2.2      &   5.6      &   0.0       &                  \\
$\mu_b$ (D)  & 0.0  &  4.3      &   1.5      &  4.8     &                 \\
\hline
\hline
\end{tabular}
\tablefoot{                                                                               \tablefoottext{a}{Experimental values from \citet{Baba2009}.}
\tablefoottext{b}{Experimental values from \citet{McNaughton2018}.}
}
\end{table*}

\begin{table*}
\caption{Theoretical molecular constants of CNP isomers.} \label{cnp}
\centering
\begin{tabular}{lccccccc}
\hline
\hline
Parameter  & Phenanthrene$^a$ & 1-CNP  &  2-CNP  & 3-CNP  &  4-CNP  &  9-CNP &  9-CNP$^b$ \\
\hline
$A$ (MHz)        & 1606.71(42)     &   1226.4 &  1610.6 &1112.4& 865.4& 841.7 &       846.118706(12)  \\
$B$ (MHz)        & 549.672(87)     &    384.2 &   311.5 & 369.3& 523.4& 487.1 &    486.386496(11)   \\
$C$ (MHz)        & 409.552(54)     &    292.6 &   261.0 & 277.2& 326.1& 308.5 &    308.9368602(69)  \\
$\mu_a$ (D)  & 0.0  &   4.1      &    5.7      &    4.6    &    0.2   &   3.7   &           \\
$\mu_b$ (D)  & 0.1  &   2.9     &     0.1     &   2.8   &    4.4   &      3.5     &      \\
\hline
\hline
\end{tabular}
\tablefoot{                                                                               \tablefoottext{a}{Experimental values from \citet{Kowaka2012}.}
\tablefoottext{b}{Experimental values from \citet{McNaughton2018}.}
}
\normalsize
\end{table*}

\begin{table*}
\caption{Experimental molecular constants of CNF isomers from \citet{Cabezas2024}.} \label{cnf}
\centering
\begin{tabular}{lcccccc}
\hline
\hline
Parameter            & Fluorene$^a$ & 1-CNF  &  2-CNF  & 3-CNF  &  4-CNF   &  9-CNF \\
\hline
$A$ (MHz)          & 2176.210153(70)  &   1303.70307(92) & 2140.700(87)   &  1541.8327(10)   & 1031.74077(27)   &  968.61817(48) \\
$B$ (MHz)          & 586.653424(70)   &   438.89904(23)  & 330.62387(37)  &  371.40436(23)   &  557.09884(28)   &  568.80591(40) \\
$C$ (MHz)          & 463.569028(20)   &   329.09001(14)  & 286.97931(34)  &  299.92528(17)   &  362.66718(23)   &  373.01098(40) \\
$\Delta_J$ (kHz)   &                  &         -        &       -        &        -         &   0.0206(18)     &   0.0359(14)   \\
$\Delta_K$ (kHz)   &                  &         -        &       -        &        -         &   $-$0.0232(21)    &        -       \\
$\chi_{aa}$ (MHz)  &                  &     0.213(72)    &          -     &    $-$2.280(39)       &    2.230(17)    &     2.229(41)   \\
$\chi_{bb}$ (MHz)  &                  &     $-$2.111(21)   &          -     &    0.379(24)        &   $-$4.137(13)    &    $-$2.339(18)   \\    \hline
\hline
\end{tabular}
\tablefoot{                                                                               \tablefoottext{a}{Experimental values from \citet{Thorwirth2007}.}\\
}
\end{table*}

\begin{table}
\small
\caption{Theoretical molecular constants of the CNAC isomers.} \label{cnacenaphthene}
\centering
\begin{tabular}{lcccc}
\hline
\hline
Parameter  &  1-CNAC  &  3-CNAC  & 4-CNAC  &  5-CNAC \\
\hline
$A$ (MHz)        & 1173.8  & 1383.4  &  1362.9  &  1213.2 \\
$B$ (MHz)        &  661.6  &  603.0  &   555.3  &   662.7 \\
$C$ (MHz)        &  441.0  &  422.1  &   396.5  &   430.8 \\
\hline
\hline
\end{tabular}
\normalsize
\end{table}

\begin{table*}
\small
\caption{Theoretical molecular constants of the DCN isomers.} \label{dcn}
\centering
\begin{tabular}{lcccccccccc}
\hline
\hline
Parameter    & 1,2-DCN  &  1,3-DCN  & 1,4-DCN  &  1,5-DCN   &  1,6-DCN & 1,7-DCN  &  1,8-DCN  & 2,3-DCN  &  2,6-DCN  &  2,7-DCN \\
\hline
$A$ (MHz)    &  1361.2  &  1004.1  & 1069.8   & 1354.3  & 1446.9       & 1111.8 &  935.6 &  1537.7   & 2620.8  & 1841.7       \\
$B$ (MHz)    &   566.7  &  600.7   &  642.6   &  556.9  &  471.3       & 571.9  &  826.1 &    476.0  &  349.2  &  381.6       \\
$C$ (MHz)    &   400.1  &  375.8   &  401.4   &  394.5  &  355.5       & 377.6  &  438.7 &    363.5  &  308.2  &  316.1       \\
$\mu_a$ (D)  & 7.0      & 4.1      &   0.0    &  0.0    &   2.1        &  4.1   &   8.6  &     8.8   &    0.0  &  0.0         \\
$\mu_b$ (D)  & 3.7      & 3.9      &   1.3    &  0.0    &   4.0        &  6.4   &   0.0  &     0.0   &    0.0  &  4.4         \\
\hline
\hline
\end{tabular}
\normalsize
\end{table*}

\section{Synthesis of cyanoacenaphthylene isomers}\label{app_synthesis}

\begin{figure*}[]
\centering
\includegraphics[width=0.9\textwidth]{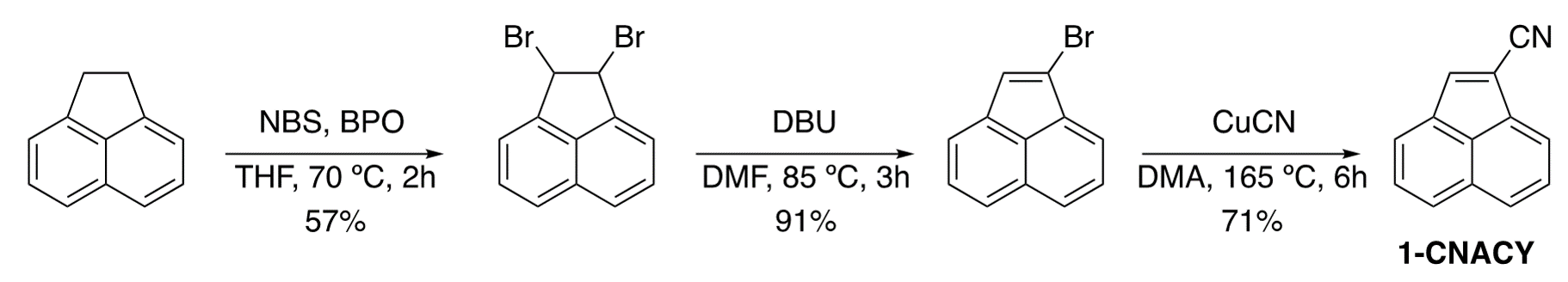}
\caption{Synthesis of 1-CNACY.}
\label{sch_1}
\end{figure*}

\begin{figure*}[]
\centering
\includegraphics[width=0.90\textwidth]{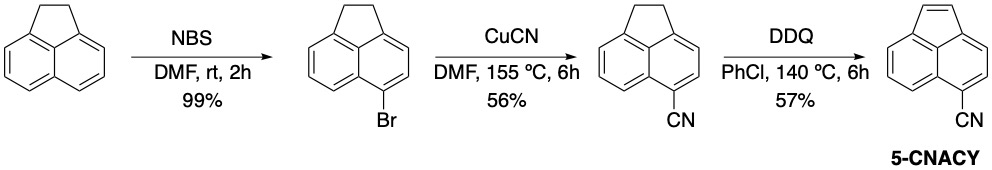}
\caption{Synthesis of 5-CNACY.}
\label{sch_2}
\end{figure*}

The synthesis of 1-CNACY and 5-CNACY was accomplished through short, three-step synthetic routes,
starting from commercially available acenaphthene (see Figs. \ref{sch_1} and \ref{sch_2}). Thus,
benzylic dibromination of acenaphthene with NBS in the presence of benzoyl peroxide afforded
1,2-dibromoacenaphthene, which after dehydrobromination with DBU and cyanodebromination with CuCN,
was converted into 1-CNACY in good yield. In contrast, treatment of acenaphthene with NBS in the
absence of the radical initiator resulted in bromination at position C5, yielding 5-bromoacenapthene
in quantitative yield. Cyanodebromination followed by oxidation with DDQ afforded 5-CNACY in
satisfactory yield.

In contrast, the 3-CNACY and 4-CNACY isomers are not accessible in a straightforward manner. In fact, they have never been prepared previously, and even the synthesis of potential precursors such as 3- or 4-bromoacenaphthylenes or -acenaphthenes would be difficult. The reason is that the reactive positions of acenaphthene are C1 (for radical reactions) and C5 (for electrophilic aromatic substitution), as shown in our synthesis. Moreover, selective functionalization of acenaphthylene in positions C3 or C4 has not been reported.

The detailed experimental procedures for the synthesis of 1-CNACY and 5-CNACY are below.

\subsection{Experimental procedures and characterization data}

1-CNACY and 5-CNACY were synthesized according to adapted literature
procedures\citep{Anderson1955,Krishnan1979,Broadus2001,Jena2020}. All reactions were
carried out under argon using oven-dried glassware. Anhydrous THF and DMF were taken from a
MBraun SPS-800 Solvent Purification System. 1-Cyanoacenaphthylene and 5-cyanoacenaphthylene
were prepared following a published procedure. Other commercial reagents were purchased from
ABCR GmbH, Sigma-Aldrich or Fluorochem, and were used without further purification. TLC was
performed on Merck silica gel 60 F$_{254}$ and chromatograms were visualized with UV light
(254 and 360 nm). Column chromatography was performed on Merck silica gel 60 (ASTM 230-400 mesh).
Centrifugation was performed in a Hettich EBA21 centrifuge. $^1$H and $^{13}$C NMR spectra were
recorded at 300 and 75 MHz (Varian Mercury-300 instrument), 400 and 101 MHz (Varian Inova 400)
or 500 and 125 MHz (Varian Inova 500) respectively. Low resolution mass spectra (EI) were obtained
at 70 eV on a HP-5988A instrument, while high-resolution mass spectra (HRMS) were obtained on a
Micromass Autospec spectrometer. Atmospheric pressure chemical ionization (APCI) HRMS were obtained
on a Bruker Microtof, using Direct Inlet Probe (DIP).

\subsubsection{Synthesis of 1,2-dibromo-1,2-dihydroacenaphthylene}

To a stirred solution of acenaphthene (500 mg, 3.24 mmol) and NBS (1.32 g, 7.45 mmol) in CHCl$_3$ (6.5 mL), benzoyl peroxide (BPO, 78.5 mg, 0.32 mmol) was added, and the mixture stirred at reflux for 2h. Solid succinimide was filtered and the filtrate was evaporated under reduced pressure. The crude product was purified by column chromatography (SiO$_2$, hexane) affording 1,2-dibromo-1,2-dihydroacenaphthylene as a yellow solid (575 mg, 57\%).
 \textbf{$^1$H-NMR} (300 MHz, CDCl$_3$), $\delta$: 7.82 (dd, $J$ = 6.7, 2.3 Hz, 2H), 7.67-7.59 (m, 4H), 6.01 (s, 2H) ppm.
 \textbf{$^{13}$C-NMR-DEPT} (75 MHz, CDCl$_3$), $\delta$: 140.51(2xC), 134.83 (C), 130.99 (C), 128.84 (2xCH), 125.86 (2xCH), 122.55 (2xCH), 54.91 (2xCH) ppm.
 \textbf{HRMS (APCI-DIP-TOF)} for C$_{12}$H$_8$Br$_2$ ([M$^+$]) Calcd.: 308.8909; Found: 309.8911.

\subsubsection{Synthesis of 1-bromoacenaphthylene}

A mixture of 1,2-dibromo-1,2-dihydroacenaphthylene (500 mg, 1.6 mmol) and DBU (292.8 mg, 1.92 mmol) in DMF (32 mL) was stirred at 85 oC for 4h. Once the mixture reached room temperature it was poured in ice water, extracted with Et$_2$O (3x25mL) and the combined organic layers were washed with 1M HCl (15mL) and saturated NaHCO$_3$ (15 mL), dried over Na$_2$SO$_4$, filtered and concentrated under reduced pressure. The crude product was purified by column chromatography (SiO$_2$, hexane) affording 1-bromonaphthylene as a brown solid (323 mg, 91\%).
\textbf{$^1$H-NMR} (300 MHz, CDCl$_3$), $\delta$: 7.87 (d, $J$ = 8.0 Hz, 1H), 7.78 (d, $J$ = 8.2 Hz, 1H), 7.70 (d, $J$ = 6.9 Hz, 1H), 7.61 (t, $J$ = 7.4 Hz, 2H), 7.53-7.47 (m, 1H), 7.15 (s, 1H) ppm.
\textbf{$^{13}$C-NMR-DEPT} (75 MHz, CDCl$_3$), $\delta$: 138.22 (C), 137.98 (C), 128.94 (C), 128.89 (CH), 128.49 (CH), 127.94 (CH), 127.67 (C), 127.63 (CH), 127.04 (CH), 123.65 (CH), 123.47 (CH), 120.73 (C) ppm.
\textbf{HRMS (APCI-DIP-TOF)} for C$_{12}$H$_8$Br ([M$^+$H$^+$]) Calcd.: 230.9804; Found: 230.9796.

\subsubsection{Synthesis of 1-CNACY.}

A mixture of 1-bromoacenaphthylene (294 mg, 1.27 mmol) and CuCN (228 mg, 2.54 mmol) in 1.3 mL of DMA was stirred at reflux for 6h. After the mixture reached room temperature, CH$_2$Cl$_2$ was added, and the resulting precipitate was filtered and washed with CH$_2$Cl$_2$. The filtrate was washed with water (15 mL) and dried over Na$_2$SO$_4$ and concentrated under reduced pressure. The crude product was purified by column chromatography (SiO$_2$, hexane/ CH$_2$Cl$_2$ 4:1) affording 1 as a red solid (146 mg, 69\%).
\textbf{$^1$H-NMR} (300 MHz, CDCl$_3$), $\delta$: 8.04 (d, $J$ = 8.2 Hz, 1H), 7.99-7.91 (m, 3H), 7.74 (s, 1H), 7.71-7.65 (m, 2H) ppm.
\textbf{$^{13}$C-NMR-DEPT} (75 MHz, CDCl$_3$), $\delta$: 139.82 (CH), 136.19 (C), 136.10 (C), 130.83 (CH), 129.23 (CH), 128.53 (CH), 128.43 (C), 128.28 (CH), 128.10 (CH), 127.23 (C), 124.71 (CH), 115.89 (C), 111.34 (C).ppm.
\textbf{HRMS (APCI-DIP-TOF)} for C$_{13}$H$_7$N ([M$^+$H$^+$]) Calcd.: 178.0651; Found: 178.0655.

\subsubsection{Synthesis of 5-bromoacenaphthene}

To a stirred solution of acenaphthene (2 g, 12.96 mmol) in dry DMF (13 mL), NBS (2.33 g, 13.13 mmol) was added in three portions and the mixture stirred at room temperature for 2 h. After that, the mixture was then transferred to a flask with ice-cold water. The precipitate was filtered and washed with H$_2$O and dried under vacuum affording 5 as s white solid
(3 g, 99\%). \textbf{$^1$H-NMR} (300 MHz, CDCl$_3$), $\delta$: 7.77 (d, $J$ = 8.4 Hz, 1H), 7.66
(d, $J$ = 7.3 Hz, 1H), 7.54 (t, $J$ = 7.8 Hz, 1H), 7.33 (d, $J$ = 7.0 Hz, 1H), 7.13 (d, $J$ = 7.5 Hz, 1H) 3.45-3.37 (m, 2H), 3.35 (m, 2H) ppm. \textbf{$^{13}$C-NMR-DEPT} (75 MHz, CDCl$_3$), $\delta$: 146.22 (C), 145.93 (C), 140.28 (C), 130.93 (C), 130.88 (CH), 129.04 (CH), 121.77 (CH), 120.16 (CH), 120.03 (CH), 116.79 (C), 30.63 (CH2), 29.91 (CH2)ppm. \textbf{HRMS (APCI-DIP-TOF)} for C$_{12}$H$_9$Br ([M$^+$H$^+$]) Calcd.: 232.9960; Found: 232.9959.

\subsubsection{Synthesis of 5-cyanoacenaphthene}

A mixture of 5-bromoacenaphthene (234 mg, 1.04 mmol) and CuCN (146.3 mg, 1.63 mmol) in DMF (3 mL) was stirred at reflux for 6h. After the mixture reached room temperature, CH$_2$Cl$_2$ was added, and the resulting precipitate was filtered and washed with CH$_2$Cl$_2$. The filtrate was washed with water (15 mL) and dried over Na$_2$SO$_4$ and concentrated under reduced pressure. The crude product was purified by column chromatography (SiO$_2$, hexane/ CH$_2$Cl$_2$ 4:1) affording
5-cyanoacenaphthene as a yellow solid (100 mg, 56\%) \textbf{$^1$H-NMR} (300 MHz, CDCl$_3$), $\delta$: 7.88 (d, $J$ = 8.3 Hz, 1H), 7.85 (d, $J$ = 7.2 Hz, 1H), 7.65 (dd, $J$ = 8.3, 7.0 Hz, 1H), 7.44-7.42 (m, 1H), 7.32 (dt, $J$ = 7.2, 0.8 Hz, 1H), 3.46 (d, $J$ = 1.0 Hz, 4H) ppm.
\textbf{$^{13}$C-NMR-DEPT} (75 MHz, CDCl$_3$), $\delta$: 153.12 (C), 147.24 (C) , 139.12 (C) , 135.05 (CH), 131.39 (C), 130.87 (CH), 121.52 (CH), 120.61 (CH), 119.27 (CH), 118.55 (C), 105.21 (C), 31.34 (CH2), 30.74 (CH2) ppm.
\textbf{HRMS (APCI-DIP-TOF)} for C$_{13}$H$_9$N ([M$^+$H$^+$]) Calcd.: 180.0257; Found: 180.0296.

\subsubsection{Synthesis of 5-CNACY.}

A mixture of 5-cyanoacenaphthene (100 mg, 0.55 mmol) and DDQ (693 mg, 3.05 mmol) in dry chlorobenzene (12 mL) was stirred at reflux for 6h. After that, the solvent was evaporated under reduced pressure and the crude product was purified by column chromatography (SiO$_2$, hexane/ CH$_2$Cl$_2$ 4:1) affording 5-cyanoacenaphthylene (5-CNACY) as a yellow solid (56 mg, 57\%). \textbf{$^1$H-NMR} (300 MHz, CDCl$_3$), $\delta$: 8.03-8.01 (m, 1H), 7.94 (d, $J$ = 7.1 Hz, 1H), 7.74-7.62 (m, 3H), 7.18 (d, $J$ = 5.3 Hz, 1H), 7.03 (d, $J$ = 5.3 Hz, 1H). ppm.  \textbf{$^{13}$C-NMR-DEPT} (75 MHz, CDCl$_3$), $\delta$: 144.70 (C), 140.25 (C), 134.82 (CH), 133.40 (CH), 130.42 (CH), 129.50 (CH), 128.52 (C), 127.78 (C), 126.32 (CH), 125.37 (CH), 123.32 (CH) , 118.03 (C), 109.54 (C) ppm. \textbf{HRMS (APCI-DIP-TOF)} for C$_{13}$H$_7$N ([M$^+$H$^+$]) Calcd.: 178.0651; Found: 178.0651.

\section{CP-FTMW spectral measurements}\label{app_FTMW}

The experimental setup employed in this work consists of a direct-digital broadband chirped-pulse Fourier-transform microwave (CPFTMW) spectrometer \citep{Neill2011} working over 2-8 GHz and located at the Universidad de Valladolid. The solid samples of 1-CNACY and 5-CNACY were vaporized at $\sim$120$^{\circ}$C in a solenoid-driven pulsed injector and diluted with an inert carrier gas (pure Ne at 2 bar). The CNACY species then expanded near adiabatically into an evacuated chamber, generating molecular jet pulses (typ. 800 $\mu$s) with effective rotational temperatures of 2 K. The molecules of each gas pulse were polarized with a series of 8 microwave chirp pulses (4 $\mu$s duration) covering the full frequency range. The measurements were performed at a repetition rate of 5 Hz, giving an effective repetition rate of 40 Hz. The chirp pulses were generated with an arbitrary waveform generator (Tektronix AWG 70002A, 25 GS/sec), amplified to 250 W with a traveling-wave tube amplifier (TWT). The excitation radiation was broadcasted perpendicularly to the jet propagation through a horn antenna. A molecular transient emission, spanning 40 $\mu$s, was then detected through a second receiving horn and amplified by a low-noise MW amplifier. A total of 1200 k FIDs per measurement were finally co-added on a digital oscilloscope (Tektronix DPO 70804C, 25 GS/sec) and a Fourier transformation finally yielded the resonance frequencies of the rotational transitions. The use of a Kaiser-Bessel apodization window produced linewidths of ca. 100 kHz. The accuracy of the frequency measurements is better than 25 kHz. All frequency components are referenced to a Rb standard.

\begin{figure*}[]
\centering
\includegraphics[width=0.90\textwidth]{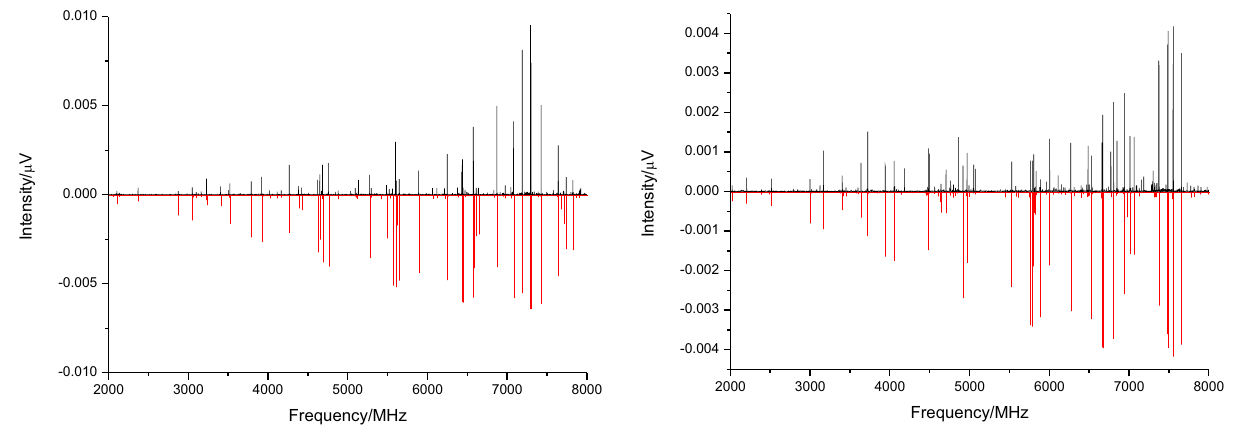}
\caption{Broadband microwave spectra of 1-CNACY (left) and 5-CNACY (right) in the region 2-8 GHz. The experimental trace is represented by the positive black trace. The negative red trace is a simulation (at 2 K) using the fitted rotational parameters of Tables \ref{1cnace} and \ref{5cnace}.}
\label{spectra}
\end{figure*}

\begin{figure*}[]
\centering
\includegraphics[width=0.90\textwidth]{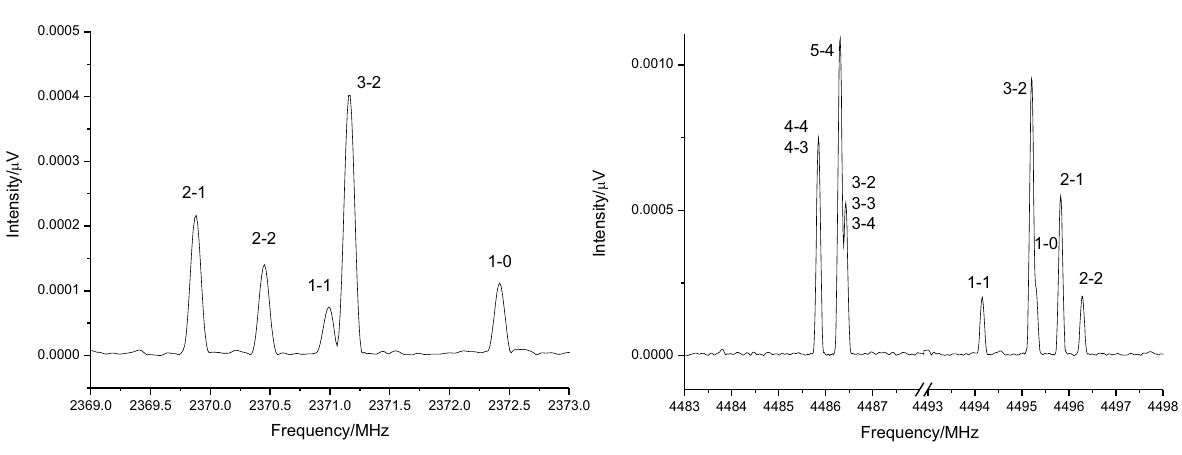}
\caption{Hyperfine spectra of the $2_{ 1,1}-1_{ 0,1}$ rotational transition of 1-CNACY (left panel) and $4_{2,3}-3_{2,2}$ and $2_{2,0}-1_{1,1}$ rotational transitions of 5-CNACY (right panel). Each hyperfine component is labelled with the corresponding values of quantum numbers $F'$-$F''$ ($F$=$I$+$J$).}
\label{hfs}
\end{figure*}

The two CNACY isomers have high $\mu_{a}$ dipole moment components, 5.6 D and 4.6 D, respectively for 1-CNACY and 5-CNACY. In addition, 5-CNACY species shows a $\mu_{b}$ dipole moment of 1.0 D. A very rich rotational spectrum is therefore observed for both isomers (see Fig. \ref{spectra}). The presence in the CNACY isomers of a $^{14}$N-nitrogen nucleus, which has a nonzero nuclear spin ($I$ = 1), causes each rotational transition to be split into several hyperfine components because of the nuclear quadrupole coupling effects (see Fig. \ref{hfs}). Therefore, the final dataset for the 1-CNACY isomer consists of 313 hyperfine components corresponding to $a$-type $R$- and $Q$-branch rotational transitions with maximum values of $J$ and $K_{a}$ quantum numbers of 12 and 5, respectively. For 5-CNACY, we measured a total of 499 hyperfine components, including $a$- and $b$-type $R$- and$Q$-branch rotational transitions with maximum values of $J$ and $K_{a}$ quantum numbers of 11 and 5, respectively.  All the observed hyperfine components were fitted using the SPFIT program \citep{Pickett1991} with the $A$-reduction of the Watson's Hamiltonian and $III^l$ representation \citep{Watson1977}. We chose this fitting method since it was employed by \citet{Thorwirth2007} to fit the spectroscopic data of the acenaphthylene molecule. A combined fit of the laboratory and TMC-1 data for each isomer provides improved values for the molecular parameters. A summary of all the obtained results is shown in Tables \ref{1cnace} and \ref{5cnace} while all the measured transitions for 1-CNACY and 5-CNACY isomers are given in Tables \ref{1cnace_ftmw} and \ref{5cnace_ftmw}. The observations corresponding to the monosubstituted $^{13}$C isotopologues will be reported separately.

The experimental molecular constants for both isomers agree very well those obtained by quantum chemical calculations. As mentioned before, we chose the $III^l$ representation. However, the values for the centrifugal distortion constants are calculated in the $III^r$ representation, and thus, the signs of the off-diagonal quartic constants, $\delta_J$ and $\delta_K$, are reversed. This fact also occurs in the case of acenaphthylene as can be seen in the data presented in Table \ref{acn_rot_constants}.

\begin{table*}
\caption{Experimental and theoretical molecular constants of 1-CNACY.} \label{1cnace}
\centering
\begin{tabular}{lccccc}
\hline
\hline
Parameter           &  Laboratory($III^l$) &   TMC-1($III^l$)     &  Lab+TMC-1($III^l$) &  Lab+TMC-1($I^r$)&Theory($III^r$)\\
\hline
$A$ (MHz)          & 1272.173002(280)$^a$ &    1272.259(46)  & 1272.173142(246)  &  1272.173107(246) &   1271.6   \\
$B$ (MHz)          &  647.282945(171)     &    647.2573(135) &  647.282793(76)   &   647.282925(77)  &    647.4   \\
$C$ (MHz)          &  429.060379(174)     &  429.061350(271) &  429.060651(60)   &   429.060554(60)  &    429.0   \\
$\chi_{aa}$ (MHz)  &    $-$4.12606(280)     &         -        &   $-$4.12641(298)   &     $-$4.12641(298) &    $-$4.37   \\
$\chi_{bb}$ (MHz)  &      2.2166(34)      &         -        &    2.2158(36)     &       2.2158(36)  &     2.44   \\
$\Delta_J$ (Hz)    &       22.33(241)     &       20.25(82)  &    21.21(133)     &         4.08(34)  &    20.44   \\
$\Delta_{JK}$ (Hz) &        9.28(225)     &       11.80(84)  &    10.61(133)     &         62.0(35)  &    11.30   \\
$\Delta_{K}$ (Hz)  &       [$-$30.04]$^b$   &        [$-$30.04]  &     [$-$30.04]      &       [$-$30.04]    &   $-$30.04   \\
$\delta_J$ (Hz)    &       $-$6.21(93)      &          [$-$6.88] &     $-$7.42(95)     &       1.144(171)  &     6.88   \\
$\delta_K$ (Hz)    &     $-$33.21(130)      &      $-$33.03(148) &    $-$32.04(44)     &       33.70(125)  &    32.23   \\
$N^c$              &         313          &         138      &       451         &           451     &            \\
$\sigma^d$ (kHz)   &         7.2          &         12.0     &        8.7        &            8.7    &            \\
\hline
\hline
\end{tabular}
\tablefoot{
\tablefoottext{a}{The uncertainties (in parentheses) are in units of the last significant digits.}
\tablefoottext{b}{Values in brackets have been kept fix to the theoretical values.}
\tablefoottext{c}{Number of transitions included in the fit.}
\tablefoottext{d}{Standard root mean square deviation of the fit.}\\
}
\end{table*}

\begin{table*}
\caption{Experimental and theoretical molecular constants of 5-CNACY.} \label{5cnace}
\centering
\begin{tabular}{lccccc}
\hline
\hline
Parameter           &  Laboratory($III^l$) &   TMC-1($III^l$)     &  Lab+TMC-1($III^l$) &  Lab+TMC-1($I^r$) &Theory($III^r$)\\
\hline
$A$ (MHz)           &   1246.694127(220)$^a$   &   1246.5655(67)    &   1246.694215(198)   &    1246.694188(197)  &    1248.9    \\
$B$ (MHz)           &   690.139405(155)   &   690.18796(258)   &    690.139409(82)    &     690.139496(82)   &     688.6    \\
$C$ (MHz)           &   444.297616(148)   &   444.29786(37)    &    444.297766(64)    &     444.297706(64)   &     443.8    \\
$\chi_{aa}$ (MHz)   &    $-$3.43615(258)    &         -          &    $-$3.43536(274)     &      $-$3.43535(274)   &     $-$3.67    \\
$\chi_{bb}$ (MHz)   &     1.52992(307)    &         -          &      1.5283(32)      &       1.5283(32)     &     1.69     \\
$\chi_{ab}$ (MHz)   &      $-$4.22(80)      &         -          &      $-$3.77(94)       &        $-$3.77(94)     &     $-$2.23    \\
$\Delta_J$ (Hz)     &      22.05(227)     &     21.12(94)      &      20.59(123)      &        9.14(42)      &     20.80    \\
$\Delta_{JK}$ (Hz)  &     $-$18.36(205)     &    $-$17.16(103)     &     $-$16.57(122)      &        17.8(35)      &    $-$16.81    \\
$\Delta_{K}$ (Hz)   &       [$-$1.97]$^b$   &      [$-$1.97]       &       [$-$1.97]        &         [$-$1.97]      &    $-$1.97      \\
$\delta_J$ (Hz)     &      $-$2.48(101)     &      [$-$3.61]       &      $-$2.18(97)       &       3.543(206)     &     3.61     \\
$\delta_K$ (Hz)     &     $-$11.85(202)     &     $-$10.22(89)     &      $-$14.31(53)      &       29.19(120)     &     12.62    \\
$N^c$              &         499          &         117        &       616            &           616     &            \\
$\sigma^d$ (kHz)   &         7.7          &         8.8        &        8.8           &            8.8    &            \\
\hline
\hline
\end{tabular}
\tablefoot{
\tablefoottext{a}{The uncertainties (in parentheses) are in units of the last significant digits.}
\tablefoottext{b}{Values in brackets have been kept fix to the theoretical values.}
\tablefoottext{c}{Number of transitions included in the fit.}
\tablefoottext{d}{Standard root mean square deviation of the fit.}\\
}
\end{table*}

\onecolumn
\begin{longtable}{cccccccccr}
\caption[]{Laboratory-observed transition frequencies for 1-CNACY$^\$$.}
\label{1cnace_ftmw}\\
\hline
\hline
 $J'$ & $K'_a$ & $K'_c$ &  $F'$  & $J''$ & $K''_a$ & $K''_c$ & $F''$ & $\nu_{obs}$  &  Obs-Calc \\
      &        &        &        &         &         &        &     &   (MHz)      &   (MHz)      \\
\hline
\endfirsthead
\caption{continued.}\\
\hline
\hline
 $J'$ & $K'_a$ & $K'_c$ &  $F'$  & $J''$ & $K''_a$ & $K''_c$ & $F''$ & $\nu_{obs}$  &  Obs-Calc \\
      &        &        &        &         &         &        &     &   (MHz)      &   (MHz)      \\
\hline
\endhead
\hline
\endfoot
\hline
\endlastfoot
\hline
  2 & 0 & 2 & 2 & 1 & 0 & 1 & 2   &       2103.611  &   $-$0.003 \\
  2 & 0 & 2 & 1 & 1 & 0 & 1 & 0   &       2103.731  &   $-$0.002 \\
  2 & 0 & 2 & 3 & 1 & 0 & 1 & 2   &       2104.897  &    0.012 \\
  2 & 0 & 2 & 1 & 1 & 0 & 1 & 1   &       2106.827  &   $-$0.001 \\
  2 & 1 & 1 & 2 & 1 & 1 & 0 & 1   &       2369.877  &   $-$0.001 \\
  2 & 1 & 1 & 2 & 1 & 1 & 0 & 2   &       2370.450  &   $-$0.000 \\
  2 & 1 & 1 & 1 & 1 & 1 & 0 & 1   &       2370.990  &    0.002 \\
  2 & 1 & 1 & 3 & 1 & 1 & 0 & 2   &       2371.162  &   $-$0.000 \\
  2 & 1 & 1 & 1 & 1 & 1 & 0 & 0   &       2372.418  &   $-$0.000 \\
  9 & 3 & 6 & 9 & 9 & 3 & 7 & 9   &       2769.378  &   $-$0.002 \\
  9 & 3 & 6 &10 & 9 & 3 & 7 &10   &       2769.593  &   $-$0.001 \\
  9 & 3 & 6 & 8 & 9 & 3 & 7 & 8   &       2769.593  &   $-$0.001 \\
  3 & 1 & 3 & 3 & 2 & 1 & 2 & 3   &       2873.588  &   $-$0.000 \\
  3 & 1 & 3 & 3 & 2 & 1 & 2 & 2   &       2874.204  &    0.003 \\
\hline
\hline
\end{longtable}
\tablefoot{
\tablefoottext{\$}{The full content of this table can be found in electronic
form at https://doi.org/10.5281/zenodo.13810127.}}
%\twocolumn

%\onecolumn
\begin{longtable}{cccccccccr}
\caption[]{Laboratory-observed transition frequencies for 5-CNACY$^\$$.}
\label{5cnace_ftmw}\\
\hline
\hline
 $J'$ & $K'_a$ & $K'_c$ &  $F'$  & $J''$ & $K''_a$ & $K''_c$ & $F''$ & $\nu_{obs}$  &  Obs-Calc \\
      &        &        &        &         &         &        &     &   (MHz)      &   (MHz)      \\
\hline
\endfirsthead
\caption{continued.}\\
\hline
\hline
 $J'$ & $K'_a$ & $K'_c$ &  $F'$  & $J''$ & $K''_a$ & $K''_c$ & $F''$ & $\nu_{obs}$  &  Obs-Calc \\
      &        &        &        &         &         &        &     &   (MHz)      &   (MHz)      \\
\hline
\endhead
\hline
\endfoot
\hline
\endlastfoot
\hline
  2 &  1 &  2 &  2 &    1 &  1 & 1 & 1  &     2022.170 &  $-$0.001 \\
  2 &  1 &  2 &  2 &    1 &  1 & 1 & 2  &     2022.628 &  $-$0.004 \\
  2 &  1 &  2 &  3 &    1 &  1 & 1 & 2  &     2023.243 &  $-$0.001 \\
  2 &  1 &  2 &  1 &    1 &  1 & 1 & 0  &     2024.266 &  $-$0.000 \\
  2 &  0 &  2 &  2 &    1 &  0 & 1 & 2  &     2202.717 &  $-$0.000 \\
  2 &  0 &  2 &  1 &    1 &  0 & 1 & 0  &     2202.844 &  $-$0.008 \\
  2 &  0 &  2 &  2 &    1 &  0 & 1 & 1  &     2203.789 &   0.012 \\
  2 &  0 &  2 &  3 &    1 &  0 & 1 & 2  &     2203.789 &   0.012 \\
  2 &  0 &  2 &  1 &    1 &  0 & 1 & 1  &     2205.419 &  $-$0.004 \\
\hline
\hline
\end{longtable}
\tablefoot{
\tablefoottext{\$}{The full content of this table can be found in electronic
form at https://doi.org/10.5281/zenodo.13810127.}}

\section{Chemical routes for the formation  of acenaphthylene}\label{chemical_routes}
Taking into account the peculiar physical conditions of TMC-1 we have to explore
barrier-less bimolecular reactions to reach the production of acenaphthylene. We describe
below three different chemical routes that could be considered in the chemical
networks for these large PAHs in cold dark clouds.

\subsection{Formation of 1- and 5-CNACY}\label{route_1}
In this route we consider that acenaphthylene is formed first and then it reacts with
cyano radicals through a barrierless addition-substitution reaction. In this case
we have the question of why only 1- and 5-CNACY are formed with similar abundances
and not the 3- and 4-isomers.

Acenaphthylene can be formed by four subroutes involving in principle
C$_{11}$+C$_{1}$, C$_{10}$+C$_{2}$, C$_{9}$+C$_{3}$ and C$_{8}$+C$_{4}$ reactions.
However, none of the C$_{8}$+C$_{4}$ reactions, C$_8$H$_8$+C$_4$H and C$_8$H$_7$+C$_4$H$_2$,
can form acenaphthylene.

The reactions C$_{11}$+C$_{1}$ are highly unlikely. Among two of these routes it is worth
to mention the reaction of atomic carbon and methyl-naphthalene (C$_{11}$H$_{10}$; no yet detected in TMC-1). However,
this reaction requires a loss of H$_2$ and an insertion of C, C$_{11}$H$_{10}$ + C $\rightarrow$ C$_{12}$H$_8$ + H$_2$.
The reaction could start by the insertion of C in the CH$_3$ group while it is known that C attacks preferentially the
aromatic rings as shown in X-beams experiments \citep[][and references therein]{Hahndorf2002}. Hence, although possible in principle, this pathway is very unlikely.

We could also consider the reaction between atomic carbon and the radical obtained by loss of an hydrogen of the CH$_3$ group of methylnaphthalene. The reaction could be C$_{11}$H$_9$ + C $\rightarrow$ C$_{12}$H$_8$ + H. As commented previously, C prefers to attack the aromatic rings and, hence, this pathway
is also very unlikely.

Finally, for the C$_{10}$+C$_2$ path the reaction of the naphthyl radical with acetylene
C$_{11}$H$_7$+C$_2$H$_2$ $\rightarrow$ C$_{12}$H$_8$ + H does form acenaphthylene,
but there is a barrier which cannot be overcome under TMC-1 conditions. However,
the reaction of naphthalene with the ethynyl radical (C$_2$H) forms on the paper acenaphthylene
(along with 1- and 2-ethynylnaphthalene), C$_{10}$H$_8$ + C$_2$H $\rightarrow$ C$_{12}$H$_8$ + H. This reaction has to be calculated. As
discussed in Sect. \ref{Discussion} is one of the most likely pathways to form the acenaphthylenes.

Among the reactions of C$_9$ + C$_3$ to consider, that of indene with C$_3$H which could form on the paper acenaphthylene. This
reaction proceeds barrierlessly via addition of CCCH either to the 6- or 5-membered rings, followed by isomerization
(ring closure, H shift) and H loss to acenaphthylene (C$_9$H$_8$ + C$_3$H $\rightarrow$ C$_{12}$H$_8$ + H). Another possible way
is the reaction of the 9-indenyl radical, C$_9$H$_7$, with the HCCCH isomer of C$_3$H$_2$.
Molecular beams experiments showed that the the 9-indenyl radical can be formed easily \citep{Yang2023}; also HCCCH – the non-detected C$_3$H$_2$ isomer – can also be formed easily as shown in crossed beams experiments \citep{Maksyutenko2011}. This pathway also proceeds barrierlessly via addition of HCCCH either to the 6 or 5 membered ring followed by isomerization (ring closure, H shift) and H loss to acenaphthylene.
These reaction has to be calculated as they are probably important formation routes of
acenaphthylene (see Sect. \ref{Discussion}).

\subsection{Formation of 5-CNACY}\label{route_2}
In this route we consider that the 5-isomer is formed from 1-cyanonaphthalene which reacts with
CCH (routes C$_{10}$ + C$_2$) to form 5-CNACY. However, the 3- and 4- isomers of CNACY could be also formed in this way. However, 1-CNACY will require additional
isomerization. In this route cyanonaphthalene is formed first. The reaction has to be calculated.

\subsection{Formation of 1-CNACY}\label{route_3}
For 1-CNACY we could consider that naphthalene is formed first and that it reacts with
CCCN leading to the formation of 1-CNACY, C$_{10}$H$_8$ + C$_3$N $\rightarrow$ 1-C$_{12}$H$_7$CN + H. This
reaction has to be calculated.

\end{appendix}

\end{document}